\documentclass[12pt,preprint]{aastex}

\usepackage{epsfig,fancyheadings}
\usepackage{rotating,longtable}

\shorttitle{Leo II}
\shortauthors{Coleman et al.}

\begin{document}

\title{A Wide-Field View of Leo II -- A Structural Analysis Using the SDSS}

\author{Matthew G.\ Coleman}
\affil{Max-Planck-Institut f\"{u}r Astronomie, K\"{o}nigstuhl 17, D-69117 Heidelberg, Germany}
\email{coleman@mpia-hd.mpg.de}

\author{Katrin Jordi}
\affil{Astronomisches Rechen-Institut, Zentrum f\"ur Astronomie der Universit\"at Heidelberg, M\"onchhofstra\ss e 12--14, D-69120 Heidelberg, Germany}

\author{Hans-Walter Rix}
\affil{Max-Planck-Institut f\"{u}r Astronomie, K\"{o}nigstuhl 17, D-69117 Heidelberg, Germany}

\author{Eva K.\ Grebel}
\affil{Astronomisches Rechen-Institut, Zentrum f\"ur Astronomie der Universit\"at Heidelberg, M\"onchhofstra\ss e 12--14, D-69120 Heidelberg, Germany}

\and

\author{Andreas Koch}
\affil{Department of Physics and Astronomy, UCLA, Los Angeles, CA, USA}

\begin{abstract}
Using SDSS I data, we have analysed the stellar distribution of the Leo II dwarf spheroidal galaxy (distance of 233 kpc) to search for evidence of tidal deformation.  The existing SDSS photometric catalogue contains gaps in regions of high stellar crowding, hence we filled the area at the centre of Leo II using the DAOPHOT algorithm applied to the SDSS images.  The combined DAOPHOT-SDSS dataset contains three-filter photometry over a $4 \times 4$ square degree region centred on Leo II.  By defining a mask in three-filter colour-magnitude space, we removed the majority of foreground field stars.  We have measured the following Leo II structural parameters: a core radius of $r_c = 2.64 \pm 0.19$ arcmin ($178 \pm 13$ pc), a tidal radius of $r_t = 9.33 \pm 0.47$ arcmin ($632 \pm 32$ pc) and a total $V$-band luminosity of $L_V = (7.4 \pm 2.0) \times 10^5 L_{\odot}$ ($M_V = -9.9 \pm 0.3$).  Our comprehensive analysis of the Leo II structure did not reveal any significant signs of tidal distortion.  The internal structure of this object contains only mild isophotal twisting.  A small overdensity was discovered appoximately $4.5$ tidal radii from the Leo II centre, however we conclude it is unlikely to be material tidally stripped from Leo II based on its stellar population, and is most likely a foreground overdensity of stars.  Our results indicate that the influence of the Galactic graviational field on the structure of Leo II has been relatively mild.  We rederived the mass-to-light ratio of this system using existing kinematic data combined with our improved structural measurements, and favour the scenario in which Leo II is strongly dominated by dark matter with $(M/L)_V \sim 100$ in solar units.

\end{abstract}


\section{Introduction}

Dwarf spheroidal (dSph) galaxies figure prominently in cold dark matter (CDM) cosmology as the lowest mass galaxies.  It is thought that clumps of dark matter formed from density fluctuations in the early Universe and underwent mergers and accretions to generate larger structures.  These small dark halos were among the sites of the very early star formation and some of the dwarf galaxies we see today are believed to be fossils of this epoch \citep{ricotti05,gnedin06}.  Indeed, the stellar population of every known dSph contains some fraction of ancient (age $\sim$ Hubble time) stars \citep{grebel97}.  These systems are strongly dominated by dark matter and kinematic data indicate that the stars reside at the centre of an extended dark halo \citep{mateo91,kleyna01,wilkinson04,koch07a,koch07c,gilmore07}.  CDM has been successful at reproducing large-scale structure in the observable Universe, however it makes some apparently discrepant predictions at the scale of dwarf galaxies \citep{moore99,freemanbh02}.

An integral part of the hierarchical merging process is the tidal disruption of small satellite halos within a large galaxy halo.  It is a matter of ongoing debate to what extent these processes have shaped the appearance, or even the survival, of the observable dSph satellites around our Milky Way.  The gravitational potential of the Galaxy is altering the structure of its orbiting dSph systems; however, the influence of the underlying dark halo of the dwarf galaxy is an unknown quantity in this process.  Consequently, the tidal distortion of satellite galaxies has recently been of interest, prompted by the discovery of the merging Sagittarius dSph \citep{ibata94,majewski03}, the Monoceros stream \citep{yanny03,newberg02} and the Canis Major object \citep{martin04}.  These distortions can also be used to measure the geometry and mass of the host and satellite dark halos (for example, \citealt{sackett94,johnston99,ibata01}) and are therefore of great import in several aspects of CDM studies.  For the present paper, we have used the Sloan Digital Sky Survey (SDSS) data to examine the global structure of the Leo II dSph also to search for tidal distortion by the Milky Way.

Leo II is a distant dSph ($d = 233$ kpc; \citealt{bell05}) of intermediate luminosity ($M_V = -10.1$; \citealt{grebel03}).  \citet{ih95} examined the (then known) eight dSph companions of the Milky Way using the Palomar and UK Schmidt photographic plate surveys, and concluded that ``the majority do not seem to be in dynamical equilibrium in their outer parts''.  Their radial profile and isopleth map of Leo II contained moderate signs of tidal distortion, however there are no overt extra-tidal structures.  However, they noted difficulties when studying Leo I and II due to their larger distances.  The best wide-field survey of this object is that by \citet{komi07} based on wide-field photometry to a depth of $V \sim 26$.  They detected a small globular cluster-like object in the halo, however there was no indication of tidal distortion.  All the dSphs have experienced some level of tidal moulding, however, given the large Galactocentric distance of Leo II, strong tidal distortions are not expected unless its orbit is highly elliptical.

\citet{aaronson85} detected carbon stars in Leo II and inferred that it contains a significant intermediate-age population.  This was confirmed by the HST study of \citet{mr96}, who found that the majority of Leo II stars were formed $9 \pm 1$ Gyr ago.  There is also an old population  of $14 \pm 1$ Gyr (a trait which appears common to all dSphs; see \citealt{grebel04}), and \citet{koch07b} found star formation in Leo II to have continued until approximately 2 Gyr ago.  \citet{bell05} found that Leo II displays a mild population gradient, with its RHB stars preferentially located at the centre of the system (another characteristic shared by other dSphs; \citealt{harbeck01}).  Indeed, \citet{komi07} have detected a small population of stars with an age of $\sim$4 Gyr located towards the centre of Leo II.  \citet{koch07b} did not detect a corresponding gradient in metallicity.  \citet{vogt95} measured the radial velocities of 31 Leo II red giants in the core region of the dSph; they measured a velocity dispersion of $6.7 \pm 1.1$ km s$^{-1}$, yielding a global $V$-band mass-to-light ratio of $11.1 \pm 3.8$ in solar units\footnote{The oft-travelled route from a {\em measured} radial velocity dispersion to a {\em calculated} mass (and hence mass-to-light ratio) requires a set of assumptions which carry considerable implications for the object at hand.  For these values \citet{vogt95} assumed the stellar motions within Leo II to be isotropic and dynamically near-equilibrium, and they also assumed that mass follows light.  We discuss this last point in \S \ref{mass} of this paper.}.  They concluded that the large velocity dispersion cannot be produced by tidal heating, thus Leo II contains an extensive dark matter halo ($\sim$$10^7 M_{\odot}$).  This result is supported by the extensive kinematic survey of \citet{koch07c}.  Their analysis of the radial velocities of 171 Leo II members over the entire luminous system found a velocity dispersion of $6.6 \pm 0.7$ km s$^{-1}$, which is essntially flat to the tidal radius, yielding a mass-to-light ratio of $27-45$ in solar units.

Here we present a wide-field photometric analysis of the Leo II dSph based on data from the Sloan Digital Sky Survey (SDSS; \citealt{york00}) Data Release 5 (DR5; \citealt{dr5}).  These data have been used to search the structure of this system for evidence of tidal distortion, and we have rederived many of the Leo II structural parameters.  SDSS is an imaging and spectroscopic survey that has mapped $\sim$$1/4$ of the sky. Imaging data are produced simultaneously in five photometric bands, namely $u$, $g$, $r$, $i$, and $z$~\citep{Fu96,Gu98,Ho01,Gu06}. The data are processed through pipelines to measure photometric and astrometric properties \citep{Lu99,stoughton02,Sm02,Pi03,Iv04,tucker06}.  SDSS has proved to be extremely useful for studying the Galactic satellites.  It has led to the discovery of tidal tails from globular clusters (for example, Palomar 5 by \citealt{oden01a}; NGC 5466 by \citealt{belokurov06a,grillmair06a}) and streams in the Galactic Halo \citep{newberg03,belokurov06b,grillmair06b,grillmair06c}.  Also, SDSS has led to the detection of several new dSph candidates \citep{willman05,belokurov06c,zucker06a,zucker06b}.  Thus, the multiple filters of SDSS are particularly useful for removing undesired sources (such as foreground stars) in the search for substructures.  This was demonstrated by \citet{oden01b} in their study of the Draco dSph and \citet{smolcic07} when examining Leo I.  In this paper we present a similar analysis of the Leo II dSph using the SDSS DR5 \citep{dr5}.

\section{Photometry} \label{sec:phot}
Part of SDSS is a set of automated algorithms to extract information from the survey images.  This includes the pipeline `Photo' \citep{lupton02,stoughton02}, which has extracted five colour photometry for the vast majority of the survey area.  However, this package was not designed or optimised to analyse crowded stellar fields: Photo contains two parameters (a time limit and a maximum number of sources) which are exceeded in regions of high stellar density.  Therefore, the centres of globular clusters and nearby galaxies are not included in the resulting SDSS photometric catalogues.  \citet{smolcic07} described an automated pipeline based on the DoPHOT program to analyse crowded SDSS fields, and then applied it to a $3.55$ square degree region centred on the dSph galaxy Leo I.  Their work contained a comparison between SDSS Photo and DoPHOT measurements, and here we present a complementary analysis of the Leo II region using the DAOPHOT package to fill the vacant areas of SDSS photometry not catalogued by SDSS's Photo.

Fig.\ \ref{sdssxy} displays all stars listed in the SDSS photometric catalogue in the region of Leo II\footnote{Here we have used `clean' photometry, removing all stars with PSF-fitting errors greater than 0.2 mag.}.  Gaps can be seen at the centre of the dSph (the region within the core radius is incomplete) and at coordinates $(0.4, 1.0)$; this second gap contains the 4th magnitude star FN Leonis (72 Leo, HD 97778)\footnote{The Bright Star Catalogue; http://www.alcyone.de/SIT/bsc/bsc.html} which resulted in Photo failing for two frames.  To complete the photometry in these areas, we accessed the SDSS images listed in Table \ref{sdssfields} in the three filters $g$, $r$ and $i$ and measured the magnitudes of all stellar sources.  In total, photometry was derived for 45 fields.  A large overlap region allowed a comprehensive comparison between the two photometry datasets.

Stellar magnitudes were measured using the DAOPHOT \citep{stetson87} program within IRAF, using the same method as described for previous surveys of the Fornax and Sculptor dSphs \citep{coleman05a,coleman05b}.  In summary, we derived a best-fitting PSF model for each image, and measured the PSF magnitude for each star in that image.  Also, we performed an aperture correction for each image by selecting stars in an uncrowded region and comparing their aperture magnitudes to those determined using the PSF-fitting technique.  In general, the PSF magnitudes were $0.01 - 0.02$ mag fainter than the aperture magnitudes.  The magnitudes for each field were adjusted to match the aperture measurement.  We removed all stars with photometry errors greater than 0.2 mag.  The stars located between overlap regions were used to ensure a constant photometric zeropoint across the fields.  An overall photometric adjustment was then made by matching stars in common with our DAOPHOT catalogue and the SDSS catalogue.  A comparison between these matched stars (after adjustment) is shown in Fig.\ \ref{photerr}.  This produced the final DAOPHOT photometry dataset.

The stated $95\%$ completeness limits of the SDSS catalogue are $g = 22.2$, $r=22.2$ and $i=21.3$ \citep{stoughton02}.  We measured these values for the DAOPHOT photometry using artificial star tests in the central Leo II field (run 5194, rerun 5, field 361).  We placed 900 stars in this field distributed on a fixed grid using the previously derived PSF as a model and then attempted to recover these stars using the same techniques as described above.  The completeness was then measured as the fraction of artificial stars successfully photometered by DAOPHOT.  The test was repeated for artificial stars in the magnitude range $18.00, 18.25,\dots , 23.00$ for all three filters.  From this, we measured $95\%$ completeness limits of $g = 22.0$, $r=21.9$ and $i=21.4$ (all with approximate uncertainties of 0.1 mag), implying that the photometric completeness returned by DAOPHOT was not affected by the stellar crowding seen at the centre of Leo II.

We measured the photometric accuracy as the dispersion of the magnitudes returned by DAOPHOT for the artificial stars.  These are indicated by the errorbars in Fig.\ \ref{photerr}.  It is instructive to determine the photometric errors of a hypothetical `average' Leo II star (that is, a star in the centre of the RGB) at the $i$-band faint limit of $i=21.3$.  An average Leo II star at this magnitude will display brightnesses of $g \approx 22.1$, $r \approx 21.6$.  From the artificial star tests, the corresponding uncertainties at these magnitudes are $\sigma_g = 0.08$ mag, $\sigma_r = 0.09$ mag and $\sigma_i = 0.08$ mag (the formal uncertainties returned by DAOPHOT were slightly smaller, $0.07-0.08$ mag).  Thus, in our DAOPHOT dataset, an average Leo II star at the photometric limit will display an uncertainty in $c_1$ and $c_2$ (these `colour' terms are defined in \S \ref{cmdfilt}) of 0.145 magnitudes.  This agrees with the $c_2$ dispersion of Leo II stars shown in Fig.\ \ref{leoIIcmd}.  This is essentially equivalent to the uncertainties associated with an equivalent star whose photometry was measured with the SDSS Photo package.  In this case, the same hypothetical Leo II star considered above will have photometric errors of $0.08$ magnitudes in all three filters.  Hence, given that the completeness limits and photometric accuracies of the DAOPHOT and SDSS catalogues are evenly matched, the two catalogues were combined to produce a final Leo II dataset.  The existing SDSS photometry displays a slightly better accuracy than ours, therefore we used the SDSS values for stars common to both catalogues.

\subsection{Stellar Population Gradient}
Colour-magnitude diagrams for the core region of Leo II are displayed in Fig.\ \ref{cmd}.  Here we have overlayed 9 Gyr age isochrones with [Fe/H] abundances of $-1.5$ and $-2.0$, via the Johnson-Cousins to SDSS solution derived by \citet{jordi06}.  Our data cover the red giant branch of this system, with the red horizontal branch stars visible at the faint limit.  Leo II is dominated by an intermediate-age, metal-poor stellar population, and this is reflected in its CMD.

Population gradients are common in dwarf spheroidal galaxies.  These reflect the early and ongoing star formation traits of the system, and, in general, the young, metal-rich population is found to be more centrally concentrated than the older, metal-poor stars (for a review, see \citealt{harbeck01}; for more recent examples refer to \citealt{tolstoy04,batt06,koch06}).  A common indicator of population gradients is the horizontal branch, whose morphology is dependent on both age and metallicity of the stellar population.  We do not have access to this feature in Leo II as it is below the flux limit (however, see \citealt{komi07} for an analysis fo the Leo II HB).  However, we are able to examine the upper RGB with some accuracy.  The colour of the upper RGB depends sensitively on changes in metallicity (as [Fe/H] increases the RGB tip moves redward) yet it displays only a slight dependence on age.  \citet{bell05} found evidence for a population gradient in Leo II and constructed a hypothesis in which ``age would be the main driver of this gradient''.  If true, then the colour of the upper RGB should not significantly change with radius

We determined the mean $(g-i)$ colour in the upper magnitude of the Leo II red giant branch.  The dataset was divided at the core radius, with the subsequent stellar populations shown in Fig.\ \ref{popgrad}.  The average upper RGB colours were measured to be,
\begin{displaymath}
\langle g-i \rangle = \left \{
\begin{array}{lll}
1.342 \pm 0.015 & \mbox{~if ~} r \le r_c; & N=60 \\
1.372 \pm 0.025 & \mbox{~if ~} r_c < r \le r_t; & N=44 \\
\end{array} \right.
\end{displaymath}
Thus, we found only tentative evidence for a redward colour shift of $0.03$ mag in the brightest magnitude of the RGB when moving beyond the core radius, however we note that this shift is within the uncertainties quoted above.  Also, an examination of Fig.\ \ref{popgrad} reveals that there is no significant difference in the gradient of the upper RGB between the inner and outer populations.  Based on medium resolution spectroscopy of 52 RGB stars covering the entire surface area of this dSph, \citet{koch07b} do not find a clear indication of any radial gradients in Leo II, neither in metallicity nor in age.  We examined the data of \citet{bosler07}, who measured [Fe/H] metallicities for 74 red giant stars in Leo II distributed out to a radius of $4'$ from the dSph centre.  Our analysis of their data found no evidence of a metallicity gradient, in support of the \citet{koch07b} result.  These results, combined with our photometric analysis, indicate that any population gradient in Leo II is at most weak.

\subsection{CMD Filtering} \label{cmdfilt}

\citet{oden01b} demonstrated that an efficient removal of field stars can be obtained by creating `object-foreground' contrast maps in colour-magnitude space with especially fit `colours', named $c_1$ and $c_2$.  By combining the three SDSS filters, $g$, $r$ and $i$, one can construct these two principal colours starting from the $(g-r)$ vs $(r-i)$ plane.  Fig.\ \ref{cc} is a colour-colour diagram containing stars within $2r_c$ of the Leo II centre.  The parameter $c_1$ is represented by the red line and was obtained by fitting a ridge-line locus to data.  Correspondingly, $c_2$ was defined to be perpendicular to this locus.  For Leo II, the two principal colours are thus represented by:
\begin{equation} \label{princolours}
\begin{array}{c}
c_1 = 0.922(g - r) + 0.387(r - i), \\
c_2 = -0.387(g - r) + 0.922(r - i),
\end{array}
\end{equation}
\noindent which then provide a simple method for CMD filtering.  For clarity, the Leo II stars are shown in Fig.\ \ref{leoIIcmd} in both the $(c_1,i)$ and $(c_2,i)$ planes, where the CMD in the second panel is seen edge-on.  The left panel contains a clear red giant branch, and this represents the primary level of filtering -- we selected those stars within the RGB area of the CMD.  The right panel indicates that the Leo II stars are dispersed around $c_2 = 0$, where the level of dispersion is defined by photometric errors.  This provided a secondary level of filtering -- we selected those stars within $2\sigma_{c_2}$ of zero, where $\sigma_{c_2}$ is the dispersion in $c_2$ as a function of $i$ magnitude, as shown in Fig.\ \ref{leoIIcmd}.  The CMD-selection was performed only on those stars above the photometric limit at $i=21.3$.

The subsequent CMD filtering method was developed by \citet{grillmair95} based on a `signal-to-noise' determination of the desired stars compared to the field population.  \citet{oden01b} applied this method using SDSS photometry of the Draco dSph, and we followed his description when applying it to Leo II.  Essentially, we constructed a CMD density function for both the Leo II and field populations, and these two functions were then compared to determine which region of the CMD minimised the field population while maintaining a high number of Leo II stars.  These CMD functions were created by dividing the $(c_1,i)$ plane into a series of cells of dimensions 0.09 magnitudes in colour and 0.35 magnitudes in $i$.  The centre of each cell was offset from the centre of its neighbouring cell by 0.015 and 0.05 magnitudes in colour and $i$ respectively, thus ensuring the CMD functions were smooth and continuous.  Contour diagrams are shown in Fig.\ \ref{cmdfuncs} which trace the signal of the Leo II and field populations, which we will refer to as $n_c$ and $n_f$ respectively.

Using these CMD density functions, the signal in each CMD cell $(i,j)$ was determined as,
\begin{equation}
s(i,j) = \frac{n_c(i,j)-gn_f(i,j)}{\sqrt{n_c(i,j)+g^2n_f(i,j)}}.
\end{equation}
\noindent  The parameter $g$ is a scaling factor defined as the ratio of the core to field areas; for our Leo II study, $g = 2.496 \times 10^{-2}$.  The core population was drawn from the region within $2r_c$ of the Leo II centre, while the region between $8r_t$ and the survey limit defined the field population.  CMD filtration was then achieved by setting a threshold signal ($s_0$) and selecting stars located in cells with $s(i,j) \ge s_0$.  We chose a threshold signal which optimised the Leo II population (see \citealt{grillmair95}), which for this dataset was $s_0=3.76$.  The contour in Fig.\ \ref{leoIIcmd} outlines the region defined for CMD selection.  Within the $s_0$ threshold contour $21162$ stars were selected across the 16 square degree region, with 3319 of these located in the central $2r_c$ region.

\section{The Structure of Leo II}
\label{structure}
Fig.\ \ref{leoIIxy} displays the distribution of the CMD-filtered stars (to the photometric limit of the datset) across the sky.  We have included both the narrow-field ($1^{\circ} \times 1^{\circ}$) and wide-field ($4^{\circ} \times 4^{\circ}$) views of the system.  A useful method for examining the structure of Galactic satellites is provided by a stellar surface density contour diagram.  Two of these are shown in Fig.\ \ref{contour} where each star has been convolved with a Gaussian of radius $1.2'$ (that is, $r_t/8$; narrow-field view, left panel) and $4.7'$ (that is, $r_t/2$; wide-field view, right panel).  An important aspect of these diagrams is the choice of contour levels.  Here we have chosen them such that the two lowest contour levels trace stellar densities $1.5\sigma_f$ and $3\sigma_f$ above the background density.  The value of `$\sigma_f$' (or, the statistical variation in the density of field stars) is dependent on two factors: the field star density, and the width of the Gaussian.  From the extra-tidal region of Leo II we know that the CMD-filtered dataset still contains a field population of density $0.341$ stars/arcmin$^2$.  For the wide-field view we used a Gaussian of radius $4.7'$, hence this is effectively the resolution of the contour diagram.  Thus, in a region with radius $4.7'$ we would expect to find 24 field stars.  Poisson statistics thus implies a background variation of $4.9$ stars (or $21\%$), and this provides our $\sigma_f$ value for the right-hand contour diagram.  For the diagram in the left panel we have a narrower Gaussian, hence the background variation is proportionally higher ($\sigma_f=1.2$ stars, or $81\%$).

Thus, choosing the width of the Gaussian is a matter of balance between the resolution and noise level per resolution element of the contour diagram.  A wide Gaussian will include more stars within its resolving radius, hence the number statistics improve.  However, this is offset by the reduction in resolution compared to a narrower Gaussian, thereby reducing the chance of finding smaller structures.  For this reason, we have chosen two viewpoints.  The wide-field view is helpful when searching for large extra-tidal structures (such as tidal tails) while the narrow view allows an examination of the internal structure of Leo II.  Both these aspects are investigated further in this section.

\citet{komi07} have noted a small knotty structure located in the outer parts of Leo II, a result which appeared after the submission of the current publication.  This structure is visible in our contour diagram, (Fig.\ \ref{contour}, {\em left panel}) at coordinates $(0.15^{\circ}, 0.0^{\circ})$.  \citet{komi07} were unable to distinguish the stellar population of this structure from that of Leo II itself, and propose that it is either some tidally stripped material, or a small globular cluster which has merged with Leo II.  An examination of our data also found that it has the same stellar population as Leo II, and we are unable to provide further insight into the nature of this structure.

\subsection{Internal Structure}
The left panel of Fig.\ \ref{contour} shows the inner square degree of our survey region.  We measured the internal structure of Leo II by fitting a series of radially increasing ellipses to the contour diagram in Fig.\ \ref{contour} (left) using the IRAF routine {\it ellipse}.  At each semi-major axis radius this program derived the best-fitting position angle, ellipticity and central coordinates using the iterative method described by \citet{jedr87}.  This algorithm was applied to 200 `half-datasets' (that is, random data subsets with $50\%$ of all stars) to achieve bootstrap-sampled values with associated uncertainties.  The results are shown in Fig.\ \ref{lscontour}.  For the remainder of this paper we adopt the structural parameters listed in Table \ref{leoIIpars} for Leo II.  These represent the uncertainty-weighted mean of the values shown in Fig.\ \ref{lscontour} for the range $0' \le r \le 9.5'$.

A characteristic associated with tidal distortion is isophotal twisting, although isophote twists can also be projection effects of triaxial distributions.  Numerical simulations (for example, \citealt{helmi01,johnston02}) predict that a tidally distorted satellite will have structural gradients, resulting in an ellipticity and position angle (and possibly central coordinates) which are dependent on radius.  These effects have been observed in the Ursa Minor dSph which contains significant structural abnormalities, including isophotal twisting \citep{spick01,palma03}.  The structure of Ursa Minor is distinctly S-shaped and it displays some of the strongest signs of tidal distortion seen in the Galactic dSphs.  We have detected an ellipticity gradient in Leo II, however there is little evidence for significant variation in position angle or central coordinates.  This is not the same level of comprehensive structural moulding observed in Ursa Minor, and the ellipticity variation seen here can be produced by triaxiality.  Hence, the internal structure of Leo II does not indicate a strong level of tidal heating.

\begin{figure}
\plottwo{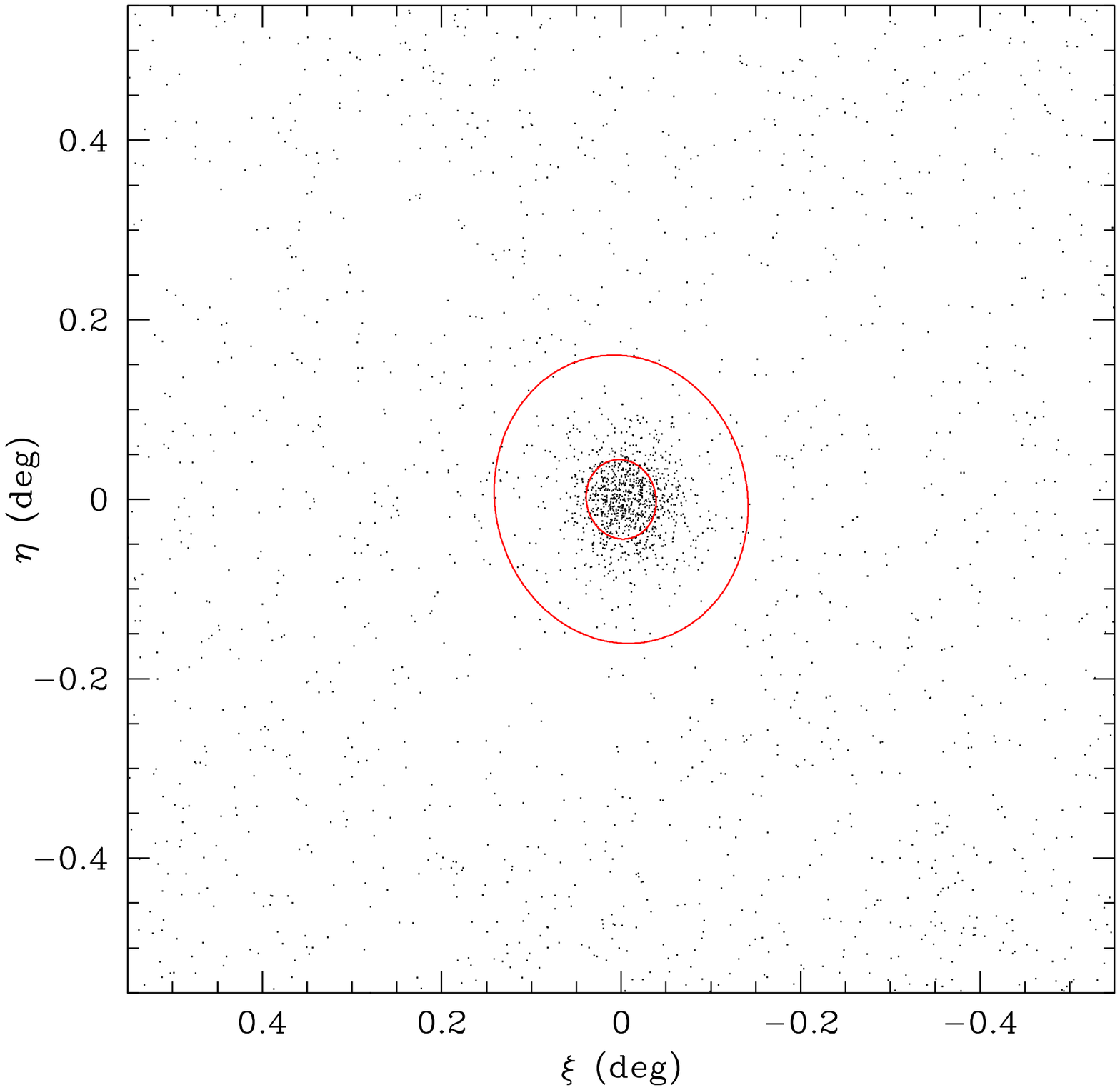}{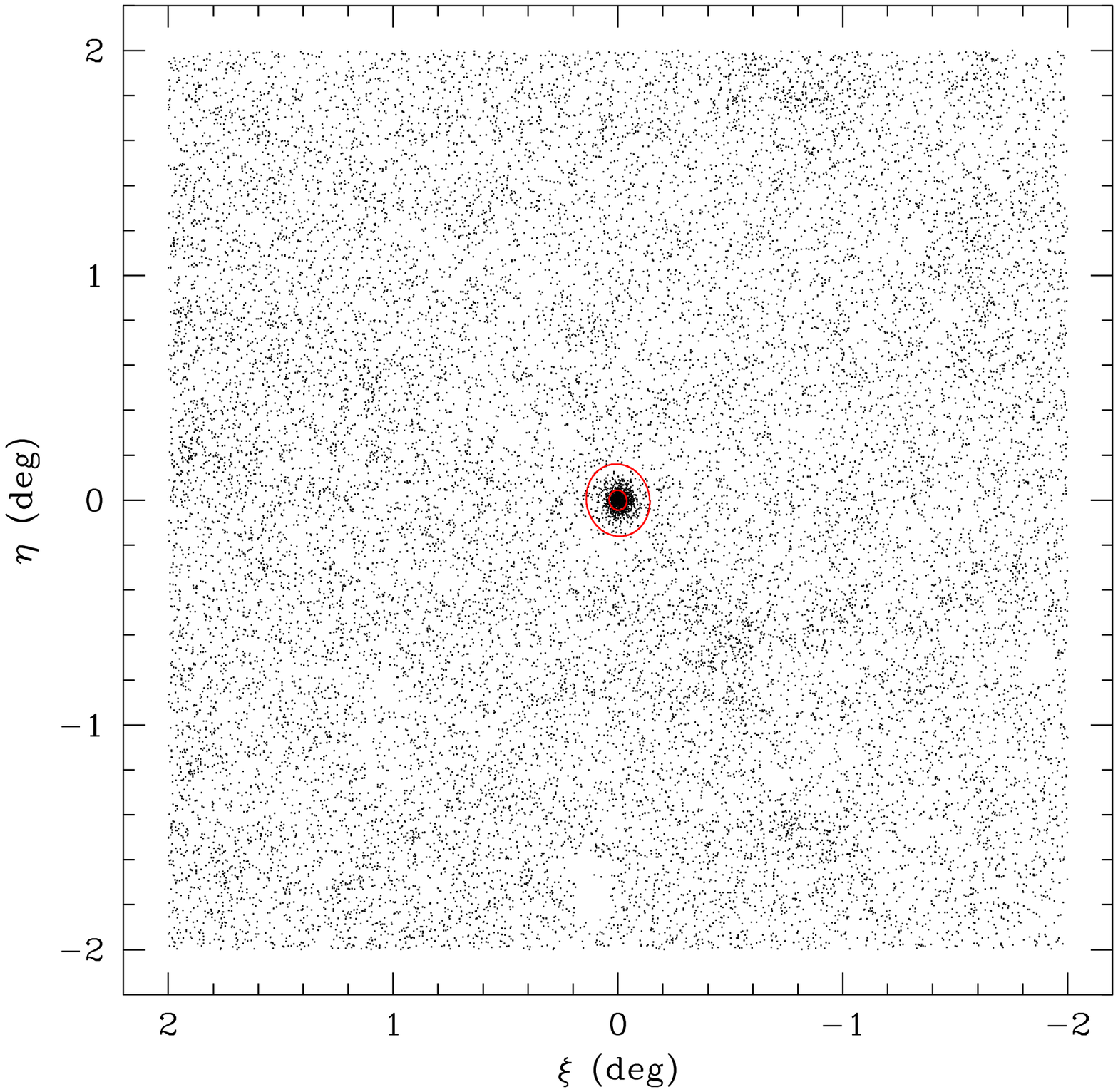}
\figcaption[Spatial distribution of CMD-selected Leo II stars]{The spatial distribution of Leo II CMD-selected stars in the narrow-field (left) and wide-field (right) views.  The core and tidal radii of Leo II are marked by red ellipses.  \label{leoIIxy}}
\end{figure}

A quantitative method of examining the structure of dSphs is provided by the radial profile, shown in Fig.\ \ref{leoIIradial}.  Here we have divided the spatial distribution of Leo II stars into a series of concentric annuli (each with an ellipticity of $e = 0.11$ and position angle of $6.7^{\circ}$) and calculated the stellar density in each.  The dashed line represents the best-fitting empirical King profile, defined by Eqn.\ 14 of \citet{king62}.  The core and tidal radii were derived using bootstrap-resampling, yielding values of $r_c = 2.64 \pm 0.19$ arcmin ($178 \pm 13$ pc) and $r_t = 9.33 \pm 0.47$ arcmin ($632 \pm 32$ pc) respectively.  Compared to \citet{komi07} ($r_c = 2.76'$, $r_t = 8.63'$) and \citet{ih95} ($r_c = 2.9' \pm 0.6$, $r_t = 8.7' \pm 0.9$), we have found Leo II to be a slightly more concentrated system of larger extent, however our values are within their uncertainties.  Our core and tidal radii are represented by the red ellipses in Fig.\ \ref{leoIIxy}.

\citet{smolcic07} examined the structure of the Leo I dSph using SDSS data and found it could also be well described by a Gaussian star density model, which allows simple modelling.  Parameters such as total luminosity are considerably simpler to calculate under a Gaussian model compared to the King model (note that a King model is essentially a truncated Gaussian) and there is little difference between the two, except for the background at large radii.  We derived a best-fitting Gaussian function to the Leo II data, and this is marked by the dotted line in Fig.\ \ref{leoIIradial}.  Using the notation of Smol\v{c}i\'{c} et al., we found a best-fitting central density of $\Sigma_0 = 30 \pm 3$ stars/arcmin$^2$ and a defining radius of $r_0 = 2.2 \pm 0.1$ arcmin.  It is worth noting that there is no significant departure of the data from either the King or Gaussian models which would support the presence of extra-tidal stars.  We found the King model to be a marginally more accurate representation of the Leo II structure, however the Gaussian model was a reasonable projection and will be used for our mass estimate of the dSph in \S \ref{lum_mass}.

\subsection{Search for Extra-Tidal Structure}

The wide-field contour diagram in Fig.\ \ref{contour} (right panel) allowed a visual search for stellar overdensities in the extra-tidal region (as defined by the King model fit).  Using this simple method, we found four structures, outlined in Fig.\ \ref{xyschem}, with stellar densities above the `$3\sigma$' level.  All these structures are beyond the range of the deep photometry by \citet{komi07}.  The highest significance structure is centred at coordinates $(-0.4,-0.6)$ and is approximately $20'$ in diameter (that is, the same apparent size as Leo II itself).  This bears some resemblance to the tidal tails seen emerging from globular clusters such as Pal 5 \citep{oden01a} and NGC 5466 \citep{belokurov06a}.  Combined with the second structure centred at $(-0.8,-1.5)$, this is qualitatively similar to the `clumpiness' seen in the Pal 5 tidal tails.  However, tidal features are generally expected (and observed) to be bi-symmetric, which is not seen here.  

If the overdensities represent material stripped from Leo II due to Galactic tidal forces, then their stellar populations should resemble that of the dSph.  The CMDs of these four extra-tidal features are displayed in Fig.\ \ref{clumpcmds}.  For comparison, we also show the data for the Leo II core and a random field region of equivalent area on the sky.  The four overdense regions have approximately twice as many stars above the completeness limit compared to the equivalent-area field region.  Here we have chosen stars within $6'$ of the centre of each overdensity and compared it to the Leo II core population defined in the previous section.  The CMDs are in shown in ($V-I$, $V$) space using the \citet{jordi06} transformation to convert SDSS magnitudes to the Johnson-Cousins filter system.  This allowed us to directly compare the data to isochrones \citep{yi01,kim02} matching the Leo II stars, a 12 Gyr age population with a metallicity between [Fe/H] $=-2.0$ and $-1.5$.

There is no obvious correlation between the Leo II stellar population and those of the overdense structures.  They all contain some stars coincident with the Leo II upper RGB, however this is not apparent at fainter magnitudes.  This is made more compelling given that, in general, the lower RGB is more densely populated than the bright end (for example, the Leo II CMD) and should therefore be more conspicuous.  To better determine if their stellar populations are coincident with that of Leo II, we statistically subtracted the field population from the CMDs of the four overdense regions.  That is, we created CMD functions for the overdensities and the field by dividing the colour-magnitude plane into a series of cells as described in the previous section, and then subtracted the field function from those of the overdense regions.  The field population was drawn from the $r > 5r_t$ region, minus the four overdense areas.  Fig.\ \ref{clumpcmdsub} shows the Hess diagrams for Leo II, the field region, and the four `field-subtracted' areas.

None of the overdense regions appear to contain a Leo II-type stellar population.  We do not believe that variations in extinction across the field could adequately shift the stellar population of a Leo II extra-tidal structure, as there is an obvious difference between the gradient of the Leo II RGB and the structures shown in Fig.\ \ref{clumpcmdsub}.  Also, a considerable change in extinction would require a small dust cloud located {\em precisely between us and the overdensities, and localised in those regions}.  This is highly unlikely given the lack of gas and extinction found in dSphs in general, and is not supported by the \citet{schlegel98} dust maps (the average reddening in the vicinity of Leo II is $E(B - V) \approx 0.019$ mag with a variation of approximately $0.004$ mag over the $4 \times 4$ square degree region).

Therefore, we conclude that it is unlikely that the overdensities consist of objects that could plausibly have been drawn from Leo II.  Moreover, the overdensities are not symmetric around Leo II, and this would argue against the tidal disruption explanation.  Thus, if the field population does not contain red giant stars at the distance of Leo II, it must be composed of foreground stars.  To coincide in CMD-space with the Leo II RGB they would have to be dwarf stars lying at distances of $1-6$ kpc.  Given the wide-field of our survey ($4 \times 4$ square degrees), it is not unreasonable to expect considerable variation in stellar density throughout the Milky Way thin and thick disk.  We believe the four overdensities are such variations.

\section{The Luminosity and Mass of Leo II}
\label{lum_mass}

\subsection{Luminosity}
We measured the total light from Leo II following two distinct methods.  Firstly, we measured the total light from Leo II in the SDSS images and subtracted the background.  From this, we measured the integrated absolute magnitude of Leo II in the SDSS filters to be $M_g = -9.67$, $M_r = -10.26$ and $M_i = -10.76$.  Using the \citet{jordi06} conversion, these correspond to $M_B = -9.25$, $M_V = -9.94$ and $M_I = -11.26$.  However, it is possible to use deep photometry of nearby globular clusters to extrapolate our photometry to fainter magnitudes.  This second method was used by \citet{smolcic07} to determine the total $I$-band luminosity of the Leo I dSph from SDSS data.

We created luminosity functions (LFs) for the core ($r \le 2r_c$) and field ($r \ge 2r_t$) regions of Leo II by counting the number of stars in bins of 0.2 mag.  The field LF was scaled by the ratio of the core-to-field spatial areas and was then subtracted from the core function.  Assuming a near-Gaussian structure, the region within two core radii is expected to contain approximately $95\%$ of the galaxy's luminosity, therefore we multiplied this differential function by a factor of $1.05$ to produce a background-subtracted Leo II luminosity function.  This function was derived in the Cousins $I$ filter to ensure it matched the globular cluster data (we made use of the SDSS to Johnson-Cousins transformations given by \citealt{jordi06}).  The resulting luminosity function is marked by the solid line in Fig.\ \ref{lf}.

To extend this luminosity function to fainter magnitudes we required a globular cluster which emulated the stellar population of Leo II.  The spectroscopic studies of \citet{bosler07} and \citet{koch07b} indicate that Leo II has a mean abundance of [Fe/H] $\approx -1.7$ with a dispersion of $0.3-0.5$ dex (all metallicities quoted in this section are on the \citealt{iti97} scale).  As such, we used the luminosity function of M5 derived by \citet{sandquist96}.  Although this cluster is more metal-rich than the mean Leo II population (the abundance of M5 is [Fe/H] $=-1.11$; \citealt{iti97}), the characteristics of the upper red giant branch are similar\footnote{We should note that M5 is a single stellar population which we are comparing to a dSph with multiple stellar populations including age and metallicity spreads (see the uncertainty analysis in the next subsection).}.  \citet{sandquist96} state their $I$-band luminosity function is $95\%$ complete to approximately one magnitude below the main sequence turnoff.  We scaled this function to match the faint end of our data and it is marked by the dashed line in Fig.\ \ref{lf}.  The remainder of the luminosity function was completed using deep HST photometry of the globular cluster NGC 6397 ([Fe/H] $= -1.95$; \citealt{harris96}).  \citet{piotto97} derived an $I$-band luminosity function for this cluster with a $90\%$ completeness to approximately eight magnitudes below the MSTO, or to stars with masses $\sim$$0.13M_{\odot}$.  This function was scaled to match the faint end of the M5 LF and is indicated by the dotted line in Fig.\ \ref{lf}.

By integrating the light from these functions\footnote{It is worth noting that our Leo II data comprised approximately one third of the total luminosity, with the remaining two thirds calculated from the scaled globular cluster LFs.} we derived a total $I$-band luminosity for Leo II of $L_I = (11.4 \pm 3.0) \times 10^5 L_{\odot}$ which is equivalent to an absolute magnitude of $M_I = -11.0 \pm 0.3$ (uncertainty analysis below).  By combining this luminosity function with the isochrones in the lower panel of Fig.\ \ref{lf} we were able to infer an integrated $(V-I)$ colour of 1.1.  This is within the uncertainty of $0.95 \pm 0.17$ measured by \citet{vogt95}.  This gave an absolute integrated $V$-band magnitude of $M_V = -9.9 \pm 0.3$ [$L_V = (7.4 \pm 2.0) \times 10^5 L_{\odot}$], which compares well with the \citet{grebel03} value of $M_V = -10.1$ and the \citet{vogt95} value of $M_V = -10.2 \pm 0.3$.  Combining these numbers with our radial profile provided the central surface brightnesses ($\Sigma_{0,V}$ and $\Sigma_{0,I}$) listed in Table \ref{leoIIpars}.  These are in agreement with the values derived by \citet{hodge82} and \citet{vogt95}.

\subsubsection{Uncertainties in the Total Luminosity}
The absolute integrated magnitude uncertainties quoted above (0.3 mag) were the product of three uncertainties.  The first of these is provided by the differences in stellar population between the globular clusters and Leo II.  Leo II is known to have a predominately red HB, yet M5 contains more blue HB stars than red.  For the $V$-band luminosity, this makes little difference, given that BHB and RHB stars are approximately the same luminosity when observed through this filter.  However, RHB stars are approximately 0.5 mag brighter than their blue counterparts in the $I$ band.  Using the luminosity function in Fig.\ \ref{lf}, we increased the $I$ band brightness of $50\%$ of the stars at the HB luminosity by 0.5 mag.  This was found to have a negligible effect, the overall brightness of Leo II changed by less than $1\%$.  Also, although the bulk of Leo II's star formation occurred between $7-14$ Gyr ago \citep{mr96}, \citet{koch07b} find evidence for the occurance of star formation until approximately 2 Gyr ago.  Thus, Leo II contains additional sub-giant and main sequence stars, which would be make it more luminous than predicted by the ancient globular cluster LFs.  From an analysis of our theoretical isochrones for Leo II combined with a Salpeter initial mass function, we estmimate this effect to have contributed approximately 0.25 mag uncertainty to the total luminosity of Leo II.

The second contribution to the uncertainty is provided by errors in our photometry.  We followed these through the LF calculation and measured their total effect on the integrated luminosity to be $0.02$ mag.  Although we do not have access to the photometry uncertainties which contributed to the luminosity functions of M5 and NGC 6397, we estimated their net influence to be approximately the same as our Leo II value.  This resulted in a combined value of 0.03 mag.

The final uncertainty stems from our scaling of the globular cluster luminosity functions to match that of Leo II.  The LFs we placed on the correct magnitude scale using their distance moduli, $(m-M)_0 = 21.84 \pm 0.13$ (Leo II; \citealt{bell05}), $14.41 \pm 0.07$ (M5; \citealt{sandquist96}) and $12.05$ (NGC 6397; \citealt{piotto97}).  The functions were then aligned by minimising the difference in number of stars ($y$-axis) in the overlapping regions.  We estimated the error in this normalisation to be approximately $10\%$, which produces a total uncertainty of $0.1$ mag in the integrated luminosity.  Combining this value with the uncertainties above in quadrature gives a total uncertainty of 0.3 mag.

\subsection{Mass} \label{mass}
An accurate determination of the total mass of a pressure-supported system requires the velocity dispersion to be measured as a function of radius.  However, the stars of a dSph galaxy presumably reside in a dark halo which dominates the mass of the system and is thought to extend far beyond the limiting radius of the luminous material (for example, \citealt{mateo91,kleyna01,wilkinson04,koch07a,pen07}).  Thus, the stars of a dSph are centrally concentrated and we are therefore able to measure the velocity dispersion only at the centre of the dark halo.  The Leo II dSph is consistent with this view: \citet{koch07c} have found that the velocity dispersion of this system is essentially constant to the luminous edge of the system.

A typical method to derive the mass of a dSph was provided by \citet{illing76}, based on the \citet{king66} models.  This method was designed for globular clusters, hence it incorporates a variety of assumptions (mass-follows-light, isotropy, dynamical equilibrium) which are inconsistent with observations of dSph galaxies.  Despite these concerns, this method provides a useful estimation for the total mass, and is still used (for example, the recent study by \citealt{simon07}).  For simple comparison to previous work, we have derived the total mass of Leo II using the Illingworth method,
\begin{equation}
M = 167 \beta r_c \sigma^2,
\end{equation}
where $r_c$ is the King core radius measured above, $\sigma$ is the velocity dispersion, and $\beta$ is a concentration-dependent parameter typically assumed to be 8 for dSphs \citep{m98}.  Using the velocity dispersion from \citet{koch07c}, we find the mass of Leo II to be $1.04^{+0.23}_{-0.21} \times 10^7 M_{\odot}$, implying a mass-to-light ratio of $14.1^{+3.4}_{-2.8}$ in solar units.  \citet{vogt95} found a similar result, deriving a `global' mass-to-light ratio of $M/L = 11.1 \pm 3.8$ in the $V$ band using their central velocity dispersion combined with an isothermal \citep{king66} model in which mass follows light.

\citet{smolcic07} outlined a different method to estimate the mass of Leo I by combining the existing kinematic data with the assumption that the total mass of the dSph structure follows a Gaussian representation within this model.  The Gaussian model is analytical \citep{rix93}, and therefore allows us to derive fundamental quantities such as the central density and todal mass of the dark halo directly from the measured parameters of Leo II.  As with other methods, this only allows a mass measurement within the {\em stellar} limiting radius.  A lower limit for the mass of Leo II can be estimated by assuming that the distribution of mass follows that of the light, while an upper limit would assume that the mass of the dark halo is effectively constant from the centre of the system to the edge of the luminous matter.  In the interests of completeness, we explore both scenarios below, however the analysis of \citet{koch07c} indicates that the upper limit is the best estimate.

From 171 radial velocity measurements, \citet{koch07c} derived a velocity dispersion for Leo II of $\sigma = 6.6 \pm 0.7$ km s$^{-1}$, which is in agreement with the previous value derived by \citet{vogt95} ($\sigma = 6.7 \pm 1.1$ km s$^{-1}$ via 31 red giants).  Given their larger dataset, we have adopted the \citet{koch07c} value in our calculations.  Assuming that the total mass (Gaussian) has a characteristic radius at least as large as the core radius (Eq.\ 10, \citealt{smolcic07}) the data then best constrain the total central mass density:
\begin{displaymath}
\rho_{0,\mbox{\scriptsize DM}} = \left( \frac{3}{4\pi} \right) \frac{\sigma^2}{Gr_0^2} = 0.13 \pm 0.04 \  M_{\odot}\mbox{ pc}^{-3},
\end{displaymath}
where the uncertainty stems from the velocity dispersion error and we have implied by our nomenclature $\rho_{0,\mbox{\scriptsize DM}}$ that this mass is dark matter dominated.  This central mass density is approximately twice the Leo I value determined by Smol\v{c}i\'{c} et al.

If we now assume for a first scenario that the total mass follows the distribution of light, this provides an estimate for the minimum mass of Leo II.  Applying Eq.\ 11 of \citet{smolcic07} we found the mass (within the tidal radius) to be $M_{\mbox{\scriptsize tot}} = 5.2^{+1.1}_{-1.0} \times 10^6 M_{\odot}$.  Using the integrated luminosity for Leo II determined earlier in the section, this produced mass-to-light ratios (in solar units) of,
\begin{displaymath}
\begin{array}{rcl}
(M/L)_V & = & 7.0^{+1.5}_{-1.4},\\
(M/L)_I & = & 4.6^{+1.0}_{-0.9},\\
\end{array}
\end{displaymath}
This scenario defines a lower limit for the Leo II mass-to-light ratio.  Hence, by comparison with the mass-to-light ratios of metal poor globular clusters, Leo II must be mostly dark matter.

However, the kinematic results of \cite{koch07c} indicate that the velocity dispersion profile of Leo II is essentially flat, and this excludes the possibility of mass-follows-light.  Therefore, we repeat the preceeding calculation assuming a large dark matter core (a model in which the central density value derived above applied to the entire luminous region), which on balance appears to be probable.  Specifically, $\rho_{\mbox {\scriptsize DM}}(r) = \rho_{0,\mbox {\scriptsize DM}}$ from the centre to the tidal radius at $r_t = 9.33'$.  Applying Eq.\ 12 of \citet{smolcic07} yielded a total mass (to the tidal limit) of $M_{\mbox{\scriptsize tot}} = 6.6^{+1.5}_{-1.3} \times 10^7 M_{\odot}$ and the corresponding mass-to-light ratios are:
\begin{displaymath}
\begin{array}{rcl}
(M/L)_V & = & 90^{+31}_{-30},\\
(M/L)_I & = & 58^{+20}_{-19}.\\
\end{array}
\end{displaymath}
This second set of values define an upper limit (we have combined the quoted uncertainties in luminosity and mass to produce the uncertainties given here).  \citet{koch07c} measured the radial velocity dispersion profile of Leo II, and their dynamical modelling excludes mass-follows-light.  This favours this second value and reinforces the trend seen in other dSphs.  The watershed study of Fornax by \citet{mateo91} found the velocity dispersion appeared to be flat with radius, indicating a dark halo with constant density to the measurement limit.  More recently, \citet{kleyna01} and \citet{wilkinson04} have found the velocity dispersions of Draco and Ursa Minor are essentially flat to the tidal radius (which is also supported by the more recent work by \citealt{munoz05}).  There has recently been several studies with similar findings for the Fornax and Sextans dSphs \citep{walker06a,walker06b}, and the Sculptor \citep{westfall06}, Carina \citep{munoz06} and Leo I \citep{koch07a} systems.  This would argue against the `mass follows light' scenario and support the second set of $M/L$ values given above.  Thus, we conclude that the mass-to-light ratio of Leo II (within its tidal radius) is $\sim$100 in solar units.

\section{Conclusion}

Our main aim was to determine whether the structure of Leo II showed signs of having been altered by the Galactic tidal field.  We have examined the structure of the Leo II dSph using SDSS photometry from DR5 supplemented with DAOPHOT photometry for the centre of Leo II.  The derived photometry was found to have an accuracy and completeness comparable to that of SDSS, thereby providing homogeneous data over wide areas.  A signal-to-noise CMD selection technique was applied to the three filter photometry to remove the majority of field stars and increase the `signal' from the Leo II dSph.

We examined the structure of Leo II in two parts, considering both its internal (to the tidal radius) and external structure.  New structural parameters were derived for this system, including its King model parameters, luminosity and mass.  These are summarised in Table \ref{leoIIpars}.  Combining our data with the kinematic study of \citet{koch07c} allowed a new determination of the Leo II mass-to-light ratio, an important parameter for dSphs given their importance in the non-linear regime of CDM.  We measured the $V$-band value to be in the range $M/L = 7 - 90$ in solar units, however we consider the upper limit to be the most likely value given that current kinematic data support a model in which the density of dark matter is constant to the radial limit of the stellar dSph system.  That is, we conclude the mass-to-light ratio in this system is $\sim$100.

We analysed the internal and external structure of this system to determine the level to which it has been disrupted by the Galactic tidal field.  The results indicated that Leo II shows some non-axisymmetry, however we did not detect any signficant evidence of extra-tidal structures such as tidal tails.  The internal structure of Leo II does not display overt distortions; the ellipticity gradient we detected could be reproduced by a triaxial structure, and there was no significant variation in the position angle or central coordinates with radius.  There was no sign of tidal distortion as seen in the Ursa Minor dSph \citep{spick01,palma03}.  In terms of regularity, Leo II resembles the Draco dSph \citep{oden01b}.

In addition, we have not detected any extra-tidal structures which can be associated with Leo II.  The only overdensities of note do not contain Leo II-type stellar populations, and it therefore does not seem plausible that they are tidally stripped material.  There are further arguments supporting this view.  Tidal tails are generally paired as a `leading' and `trailing' arm, yet the detected structures are not symmetric around Leo II.  Also, the large Galactocentric distance of Leo II would suggest a minimal level of tidal distortion.  It is essential to obtain velocity information to clarify the role of tides and to establish the nature of the overdensity (the VLT data of \citealt{koch07b} cover the surface of Leo II, however they do not extend to the region of the overdensity).

Leo II is a distant dSph, and if this is true for the entirety of its orbit, then strong tidal distortions are not expected.  \citet{vogt95} and \citet{koch07b} did not find strong distortions, and our results support the view that any tidal deformation of this system has been relatively tranquil.

\acknowledgments
The authors wish to thank the anonymous referee for their helpful comments which improved this publication.  M.C. thanks Dan Zucker, Jelte De Jong and Eric Bell for their help with SDSS in general.  K.J., A.K., and E.K.G.\ gratefully acknowledge support by the Swiss National Science Foundation through the grants 200020-105260 and 200020-113697.

Funding for the SDSS and SDSS-II has been provided by the Alfred P. Sloan Foundation, the Participating Institutions, the National Science Foundation, the U.S. Department of Energy, the National Aeronautics and Space Administration, the Japanese Monbukagakusho, the Max Planck Society, and the Higher Education Funding Council for England. The SDSS Web Site is http://www.sdss.org/.

The SDSS is managed by the Astrophysical Research Consortium for the Participating Institutions. The Participating Institutions are the American Museum of Natural History, Astrophysical Institute Potsdam, University of Basel, University of Cambridge, Case Western Reserve University, University of Chicago, Drexel University, Fermilab, the Institute for Advanced Study, the Japan Participation Group, Johns Hopkins University, the Joint Institute for Nuclear Astrophysics, the Kavli Institute for Particle Astrophysics and Cosmology, the Korean Scientist Group, the Chinese Academy of Sciences (LAMOST), Los Alamos National Laboratory, the Max-Planck-Institute for Astronomy (MPIA), the Max-Planck-Institute for Astrophysics (MPA), New Mexico State University, Ohio State University, University of Pittsburgh, University of Portsmouth, Princeton University, the United States Naval Observatory, and the University of Washington.


\cleardoublepage

\plotone{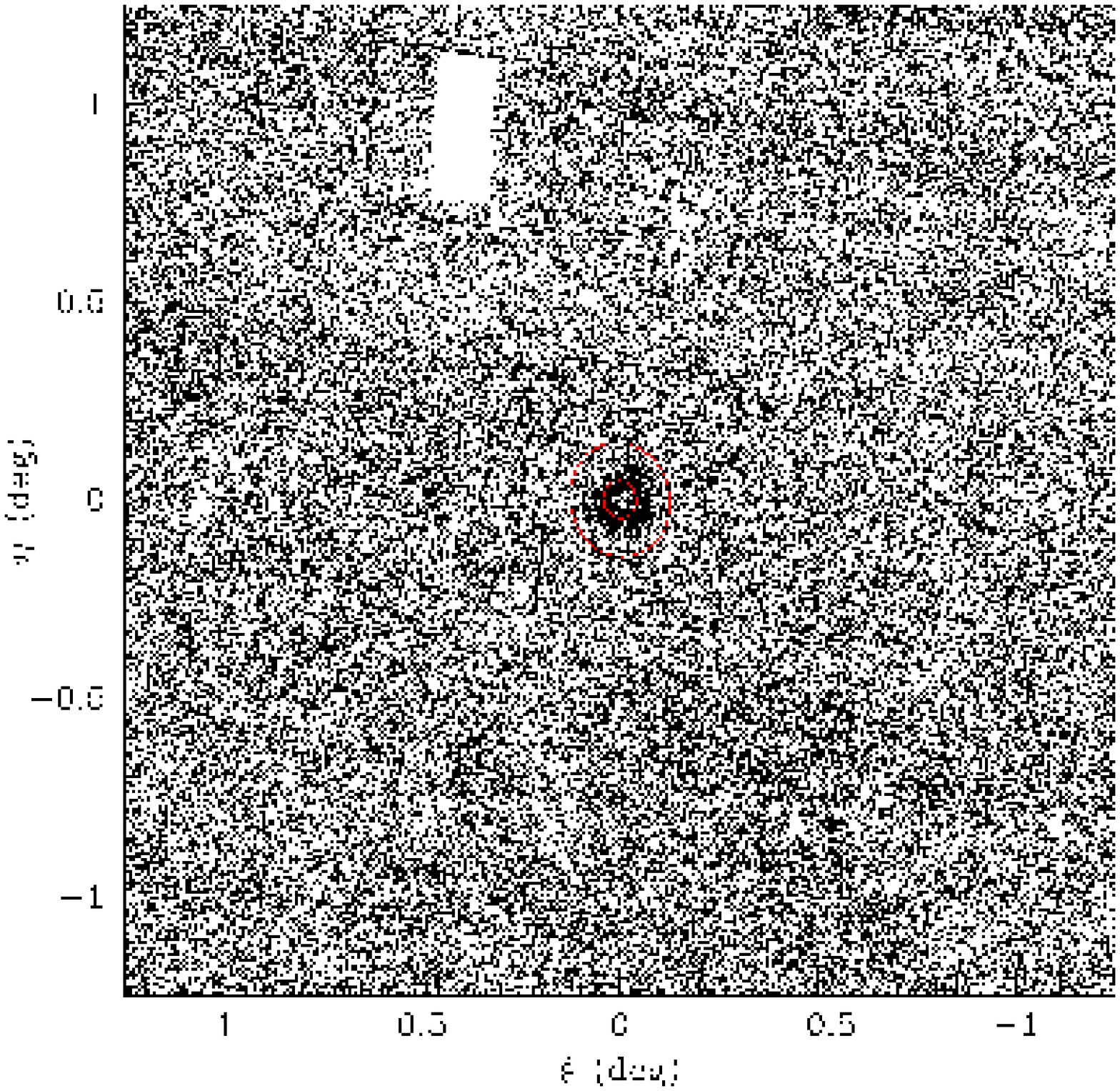}
\figcaption[Leo II SDSS photometry]{The Leo II region as it exists in the SDSS catalogue.  Note the missing stars at the centre of the dSph and towards the upper left corner.  The red ellipses trace the core and tidal radii of the Leo II dSph \citep{m98} and the dashed lines outline the region within which we extracted $g$, $r$ and $i$ photometry using the DAOPHOT package.  \label{sdssxy}}

\plotone{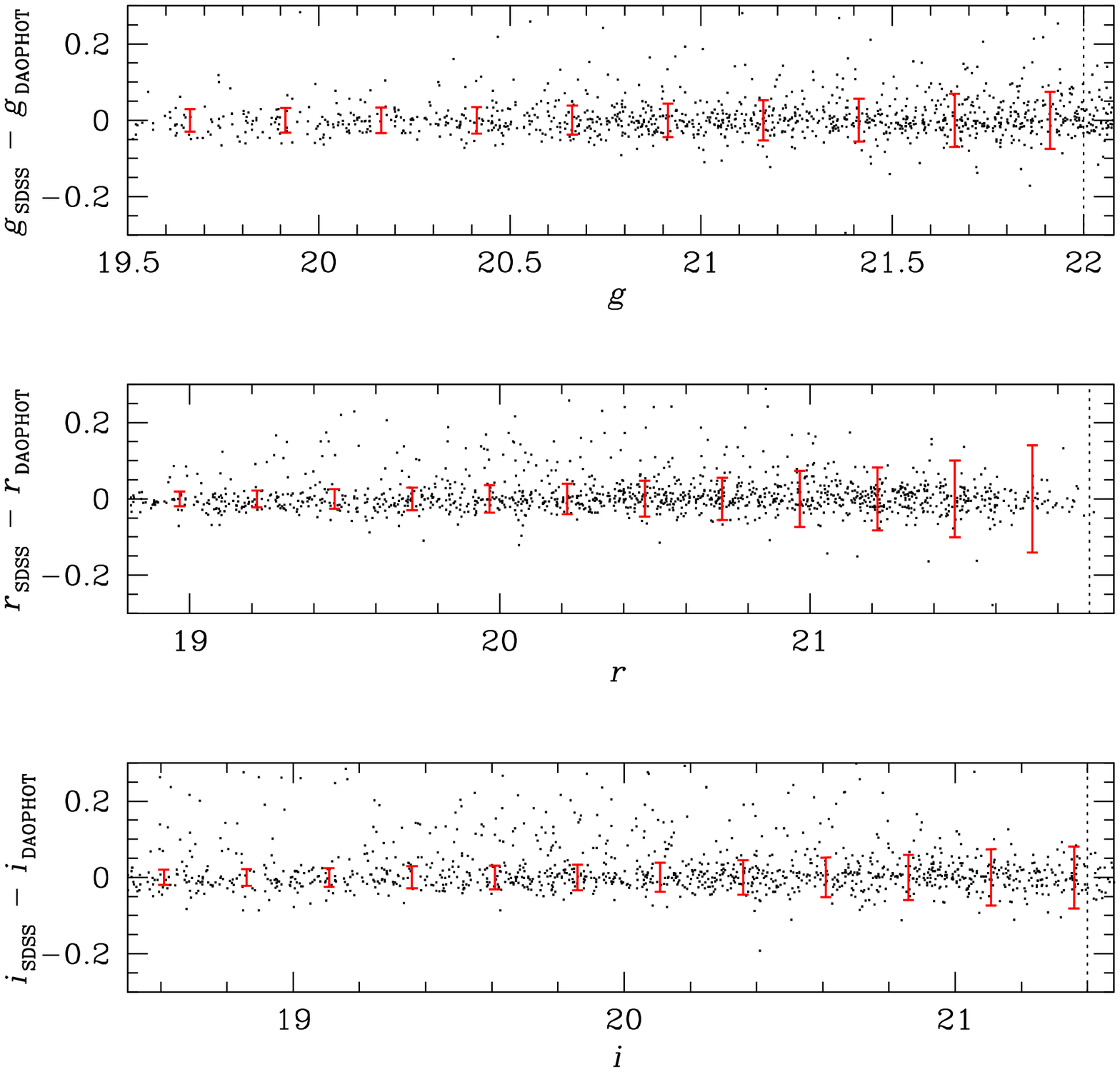}
\figcaption[Leo II photometric error]{Photometric error for the Leo II dataset as a function of magnitude.  The black points represent those stars matched between the DAOPHOT and SDSS datasets.  The uncertainty bars ($\pm 1\sigma$) come from the artificial star tests performed on the central Leo II field, and the dotted lines represent the $95\%$ completion limits returned by artificial star tests in our DAOPHOT dataset.  \label{photerr}}

\plotone{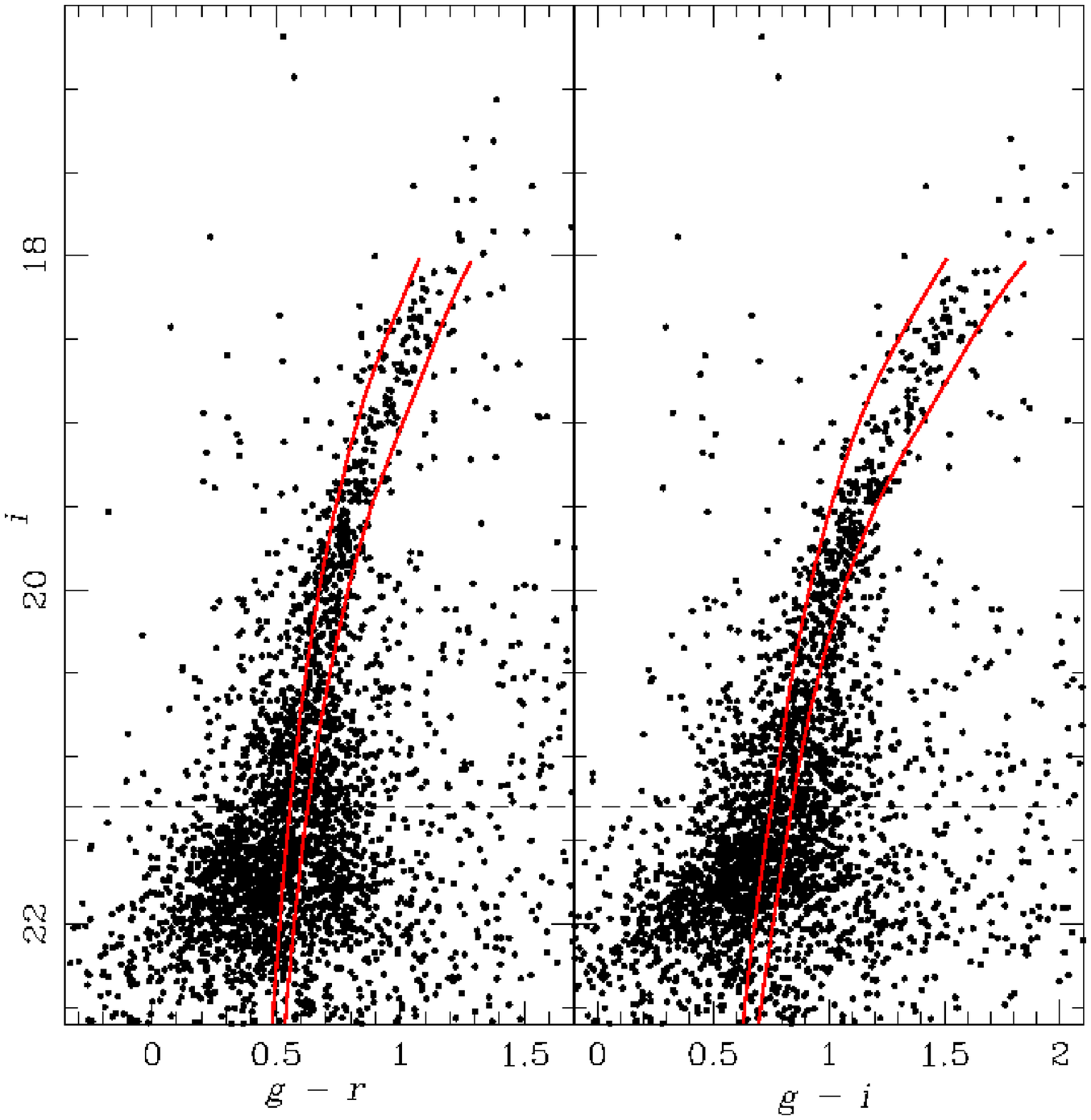}
\figcaption[Leo II CMD]{Colour-magnitude diagrams for the core region ($r \le 5.8'$) of Leo II.  The left panel contains the $(g-r)$ vs $i$ data, while the right-panel contains the $(g-i)$ vs $i$ data.  The isochrones \citep{yi01,kim02} in each panel trace a 9 Gyr age population with [Fe/H] abundances of $-1.5$ and $-2.0$, converted to the SDSS filters using the empirical solutions of \citet{jordi06}.  The dashed line represents the $i=21.3$ photometric limit of our survey.  \label{cmd}}

\plotone{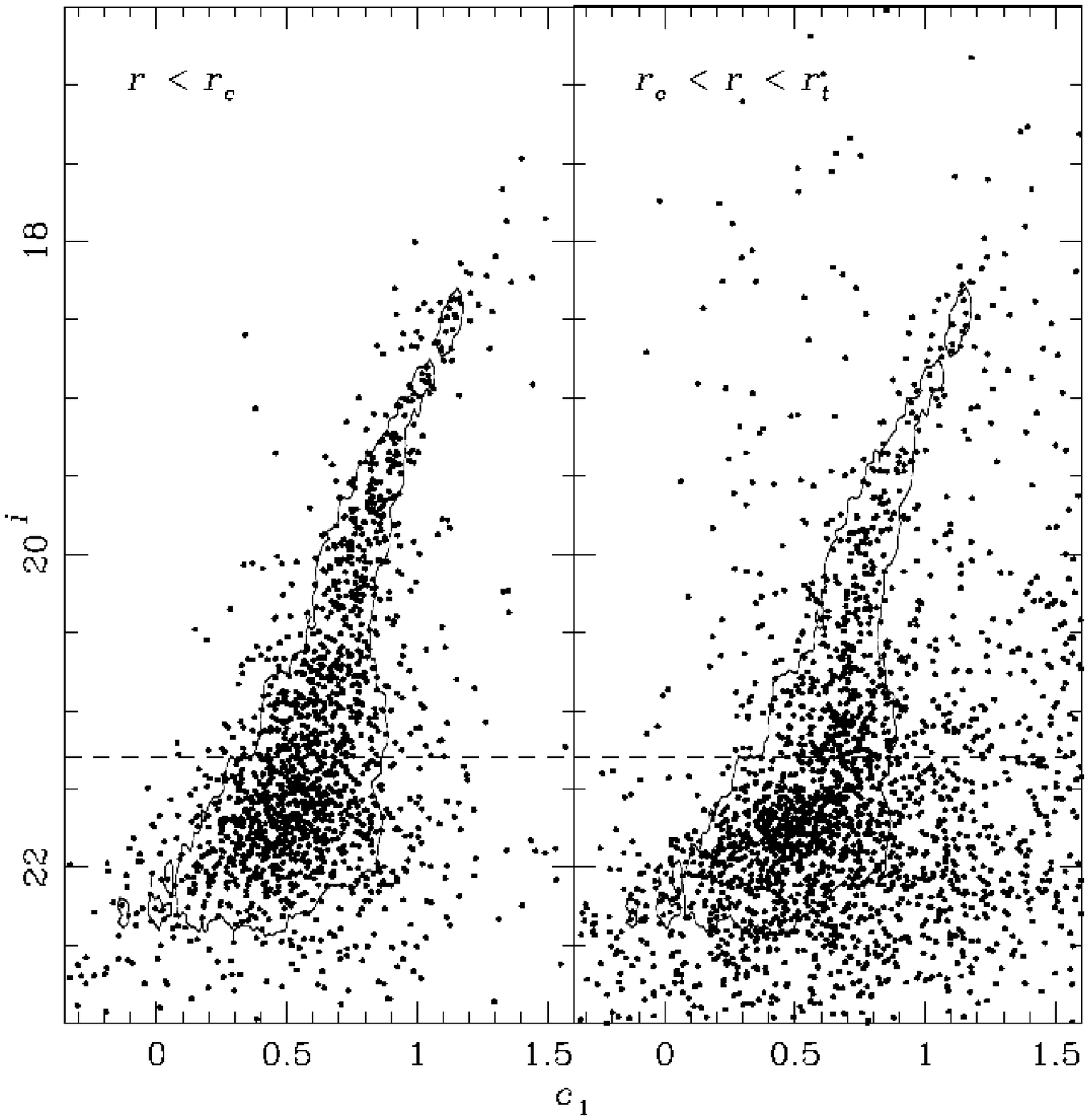}
\figcaption[Leo II population gradient]{The stellar populations of the inner ($r < r_c$) and outer ($r_c < r < r_t$) regions of Leo II overlaid on the CMD selection region from Fig.\ \ref{leoIIcmd}.  No obvious change in the upper RGB colour between the inner and outer populations is visible.  \label{popgrad}}

\plotone{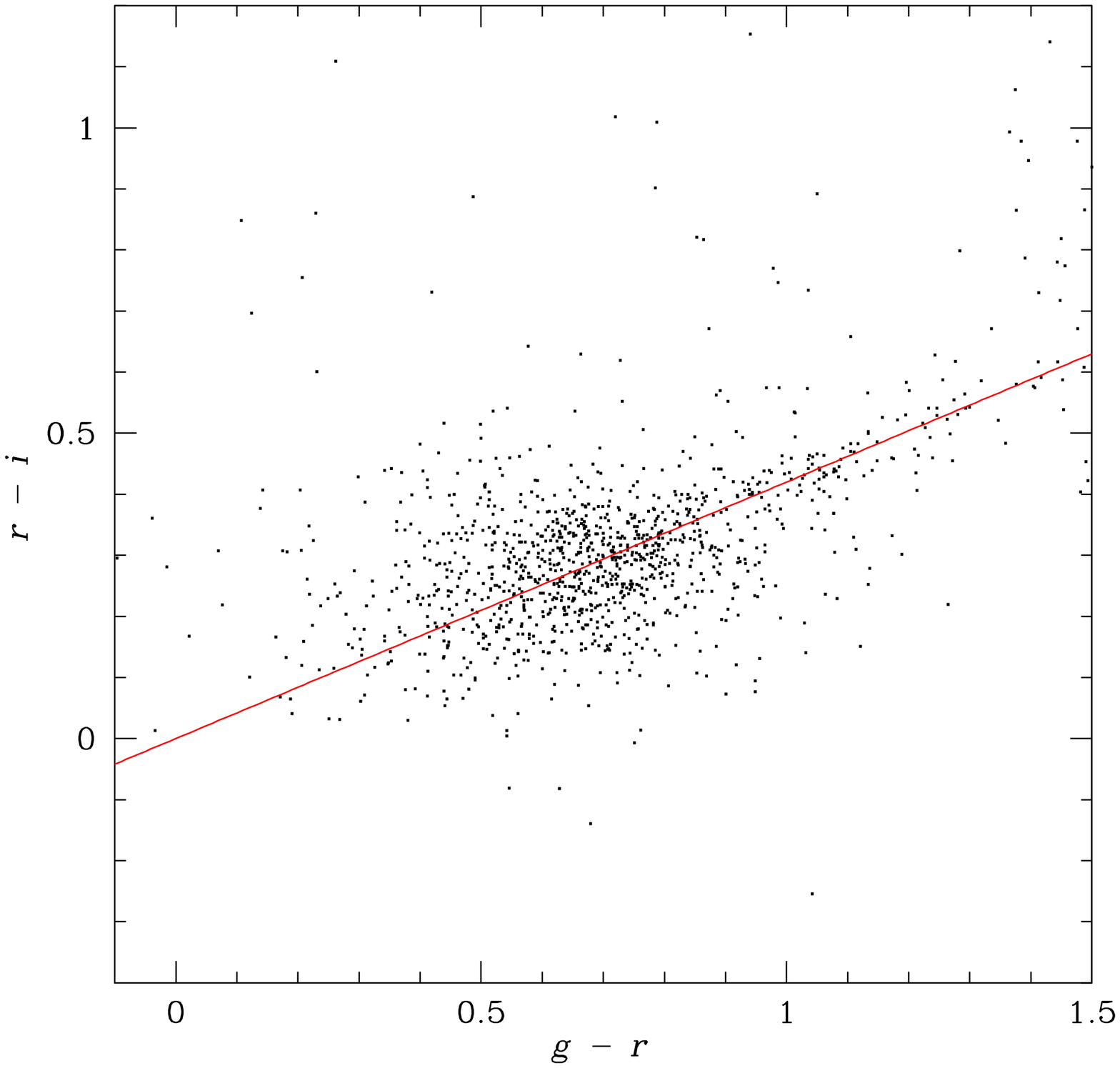}
\figcaption[Leo II colour-colour diagram]{Colour-colour diagram for stars in the central region ($r \le 2r_c$) of Leo II.  The best-fitting line shown defines the locus of $c_1$.  \label{cc}}

\plotone{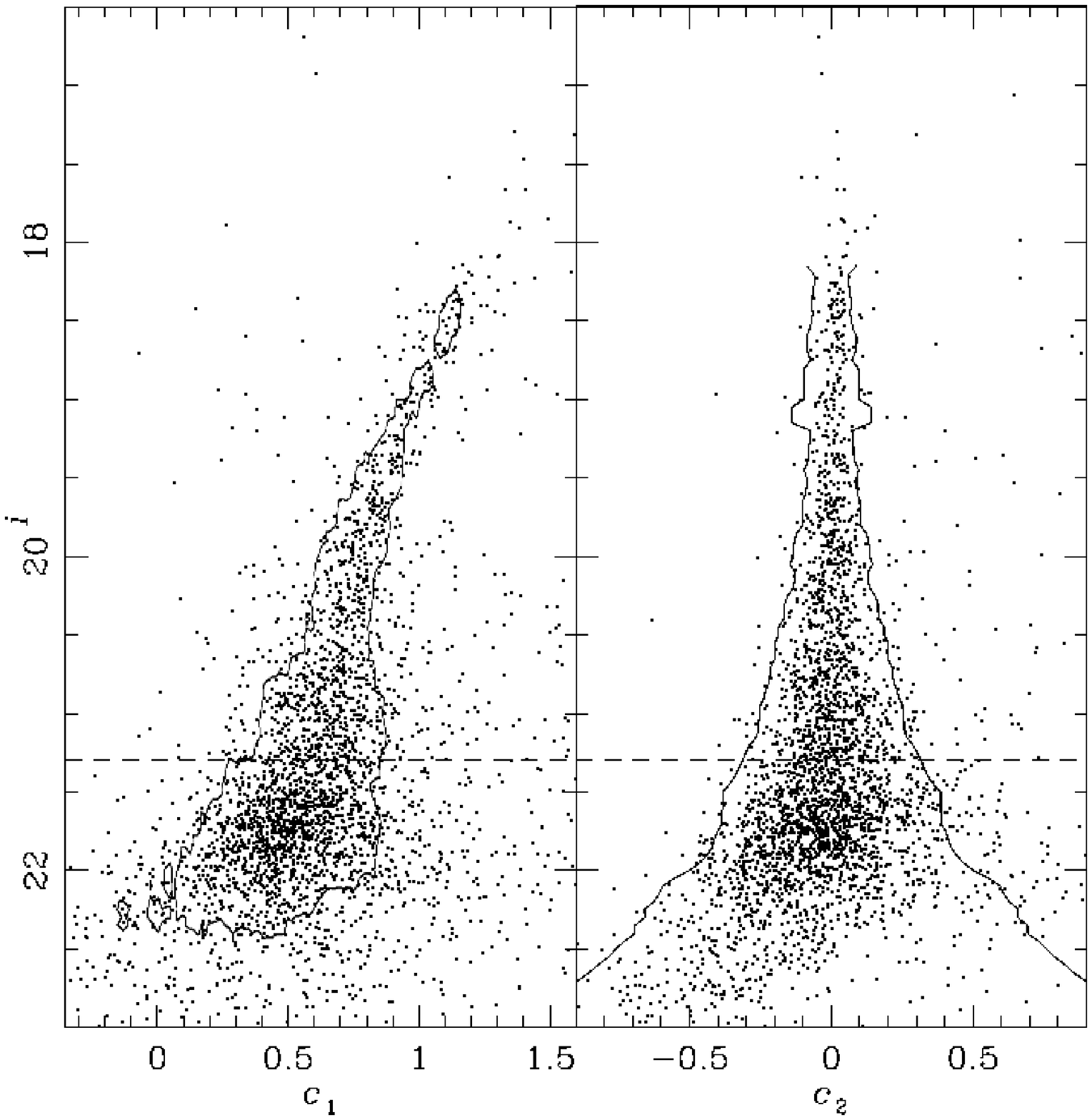}
\figcaption[Leo II CMD]{CMDs for the core region of Leo II.  The left panel shows $c_1$ vs $i$ magnitude and contains the upper red giant branch of Leo II.  The right panel shows the same stars in the ($c_2$, $i$) plane, where the dispersion of these values around $c_2 = 0$ is due to photometric uncertainties.  The dashed line represents the $i=21.3$ magnitude limit of the photometry and the solid lines trace the Leo II selection limits.  In the left panel this was defined by the comparison of the core and field populations in Fig.\ \ref{cmdfuncs}, while the selection limit in the right panel was to be $2\sigma_{c_2}$, where $\sigma_{c_2}$ is the dispersion of the central Leo II stars around $c_2=0$.  \label{leoIIcmd}}

\clearpage

\plotone{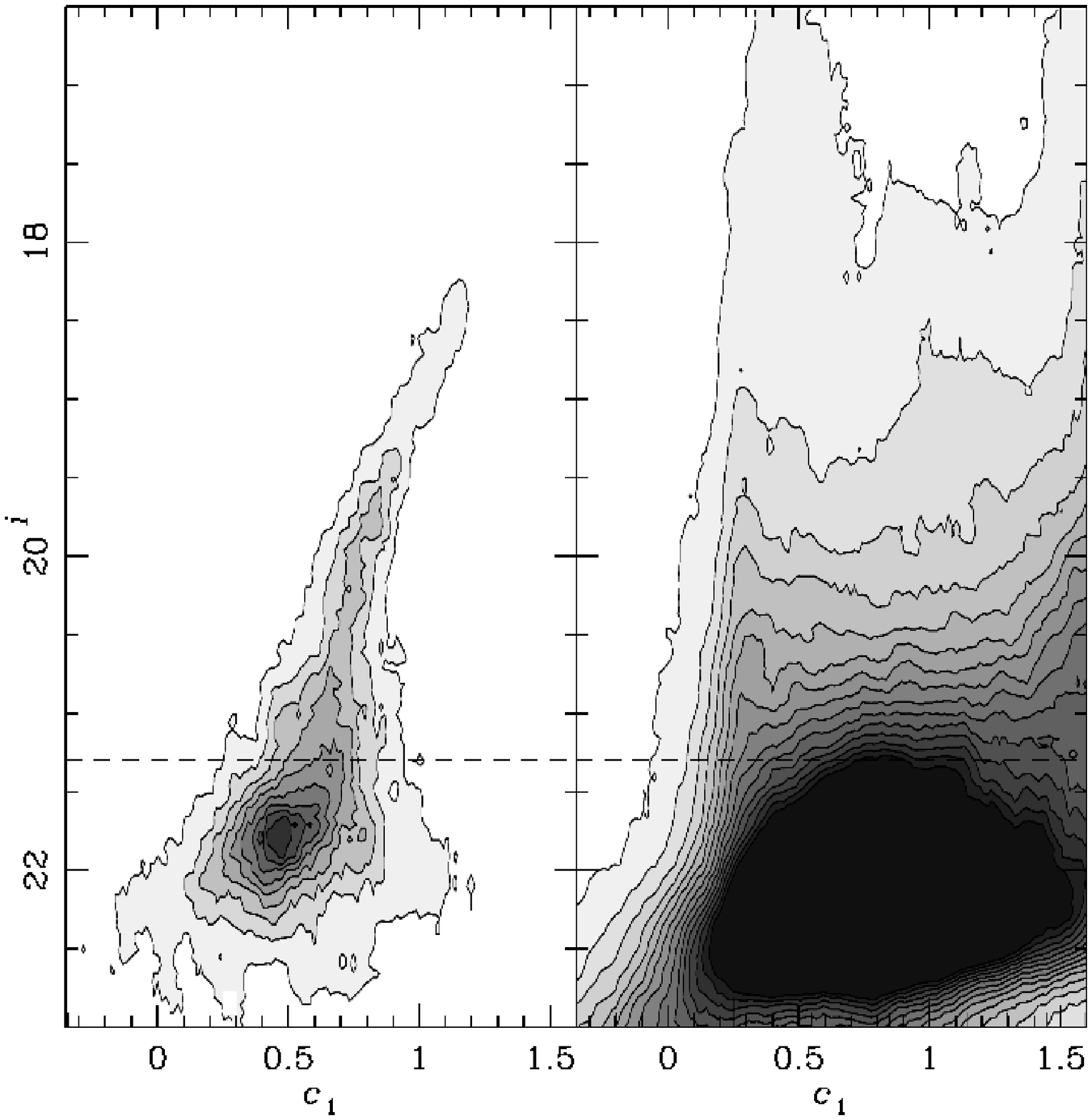}
\figcaption[Core and field CMD functions]{The CMD functions for the core (left) and field (right) populations of Leo II, where a darker colour represents a hihger stellar density.  The core population was drawn from the inner $2r_c$ region of Leo II, while the field population came from the region between $r = 8r_t$ and the survey limit.  The ratio of these two functions representing Leo II's contrast in colour-magnitude space produced the limiting signal represented in the left panel of Fig.\ \ref{leoIIcmd}.  \label{cmdfuncs}}

\clearpage

\plotone{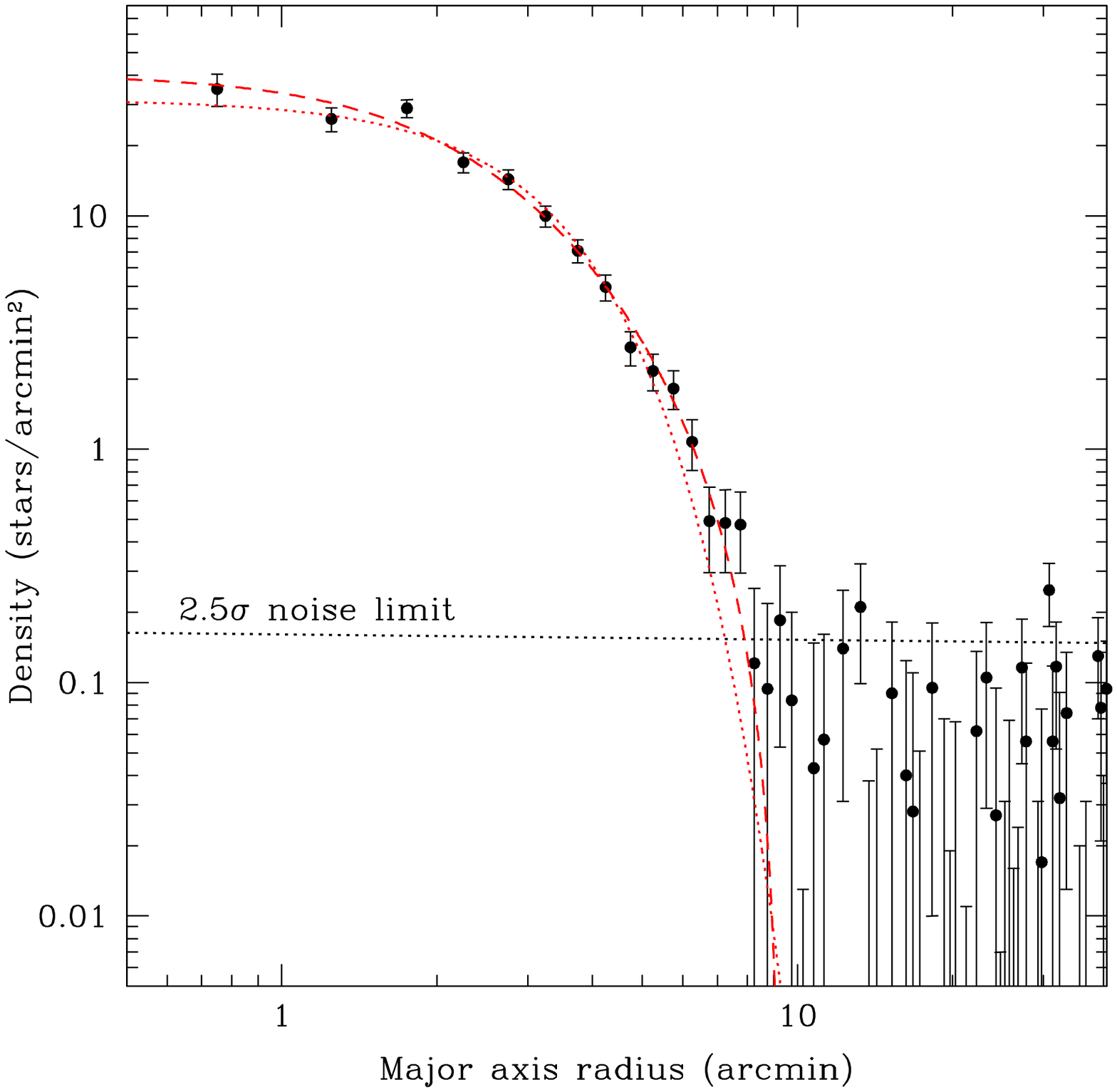}
\figcaption[Leo II radial profile]{Radial profile of Leo II, where the dashed line represents the best-fitting King profile and the dotted line traces the best-fitting Gaussian function.  A background level of $0.341$ has been subtracted from all data points.  The lower dotted line represents the $2.5\sigma$ noise limit, where $\sigma = 0.075$ stars/arcmin${}^2$ is the dispersion of the radial densities around the background level.  \label{leoIIradial}}

\begin{figure}
\plottwo{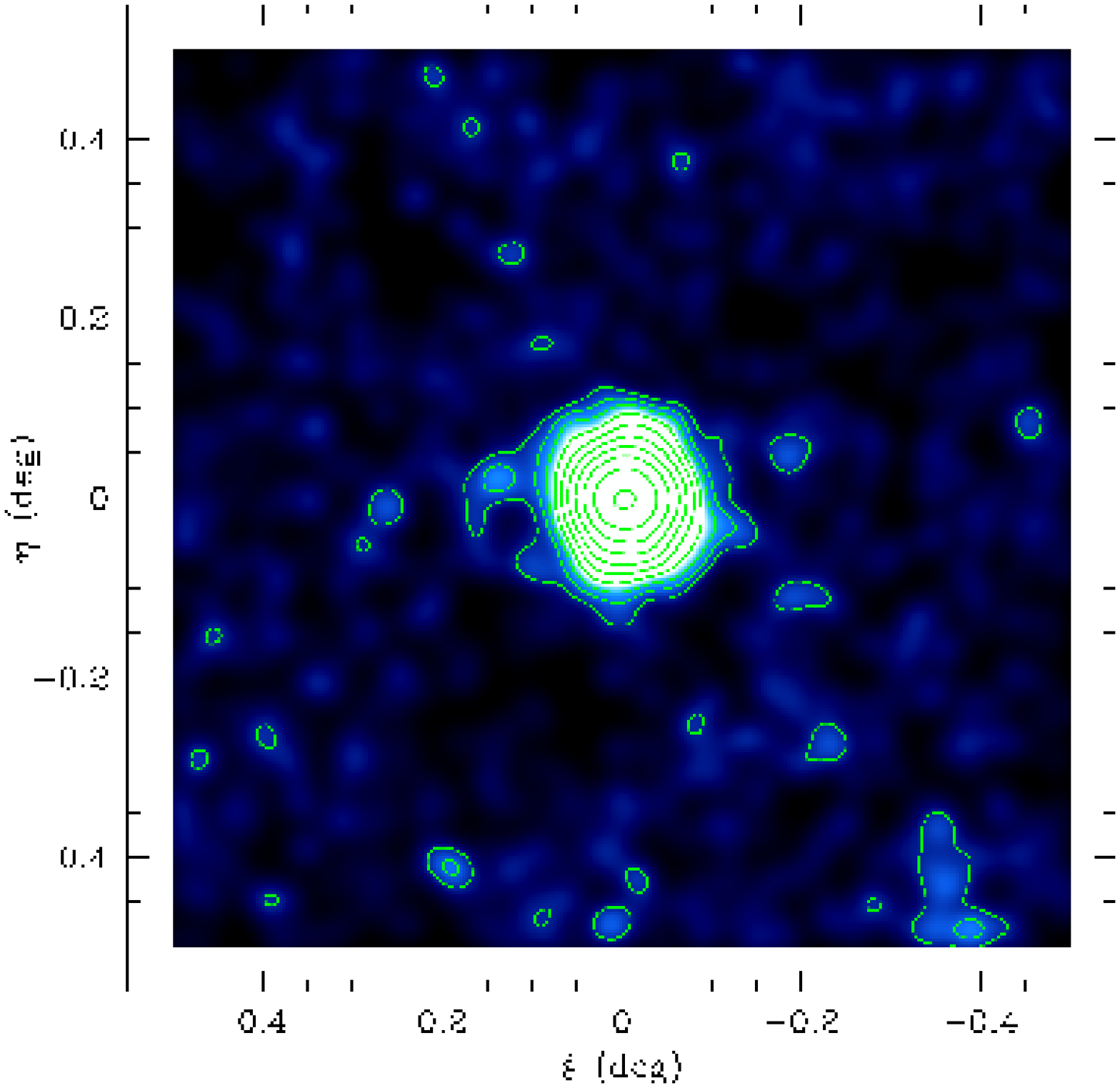}{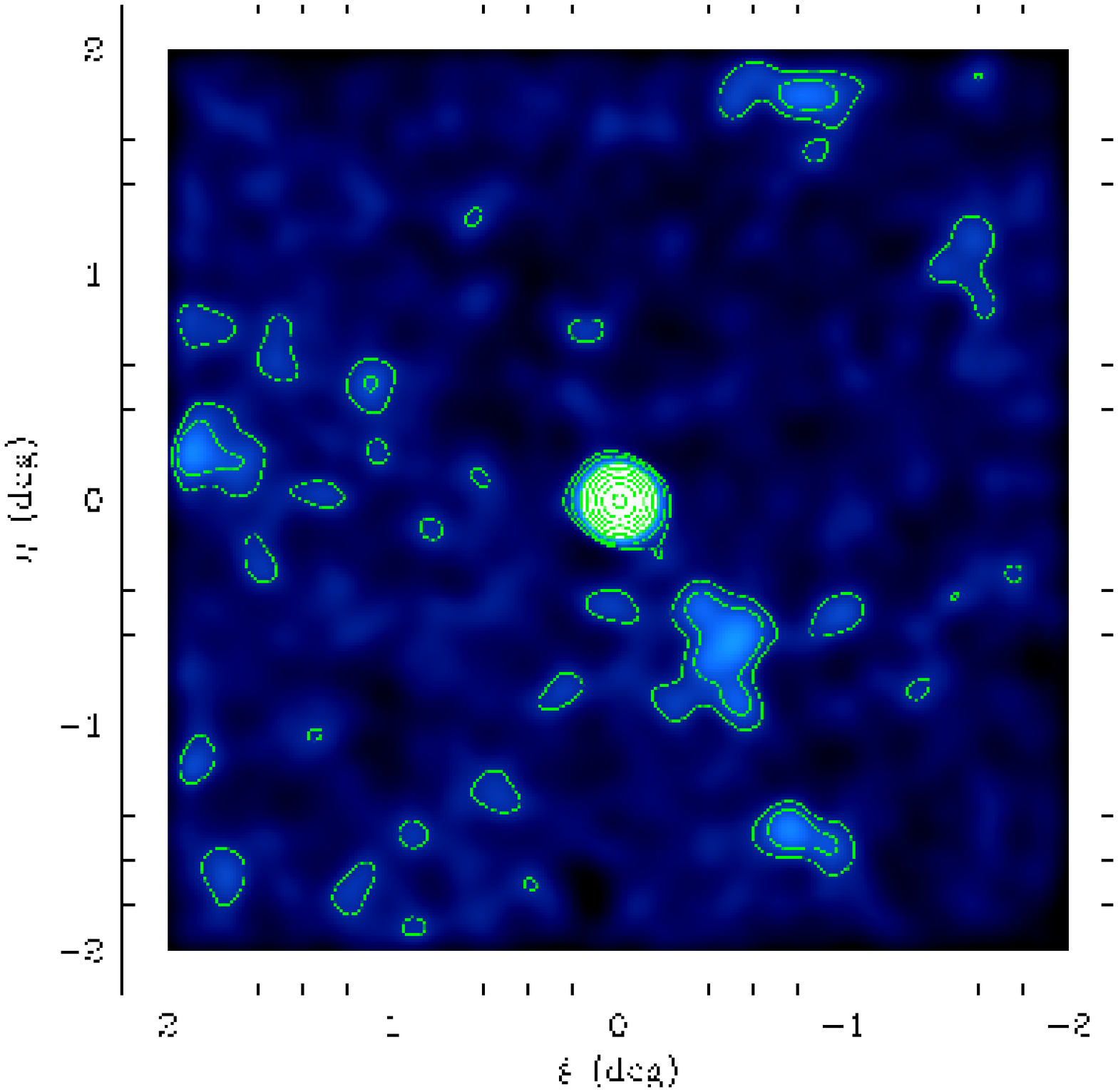}
\figcaption[Contour diagrams of Leo II]{Stellar density contour diagrams of Leo II, showing the narrow-field (left) and wide-field (right) views.  Stars in the narrow-field have been convolved with a Gaussian of radius $1.2'$, and the function in the wide-field has a radius of $4.7'$.  In both diagrams, the first and second darkest contours trace density levels above the background of $1.5\sigma$ and $3\sigma$ respectively. \label{contour}}
\end{figure}

\clearpage

\plotone{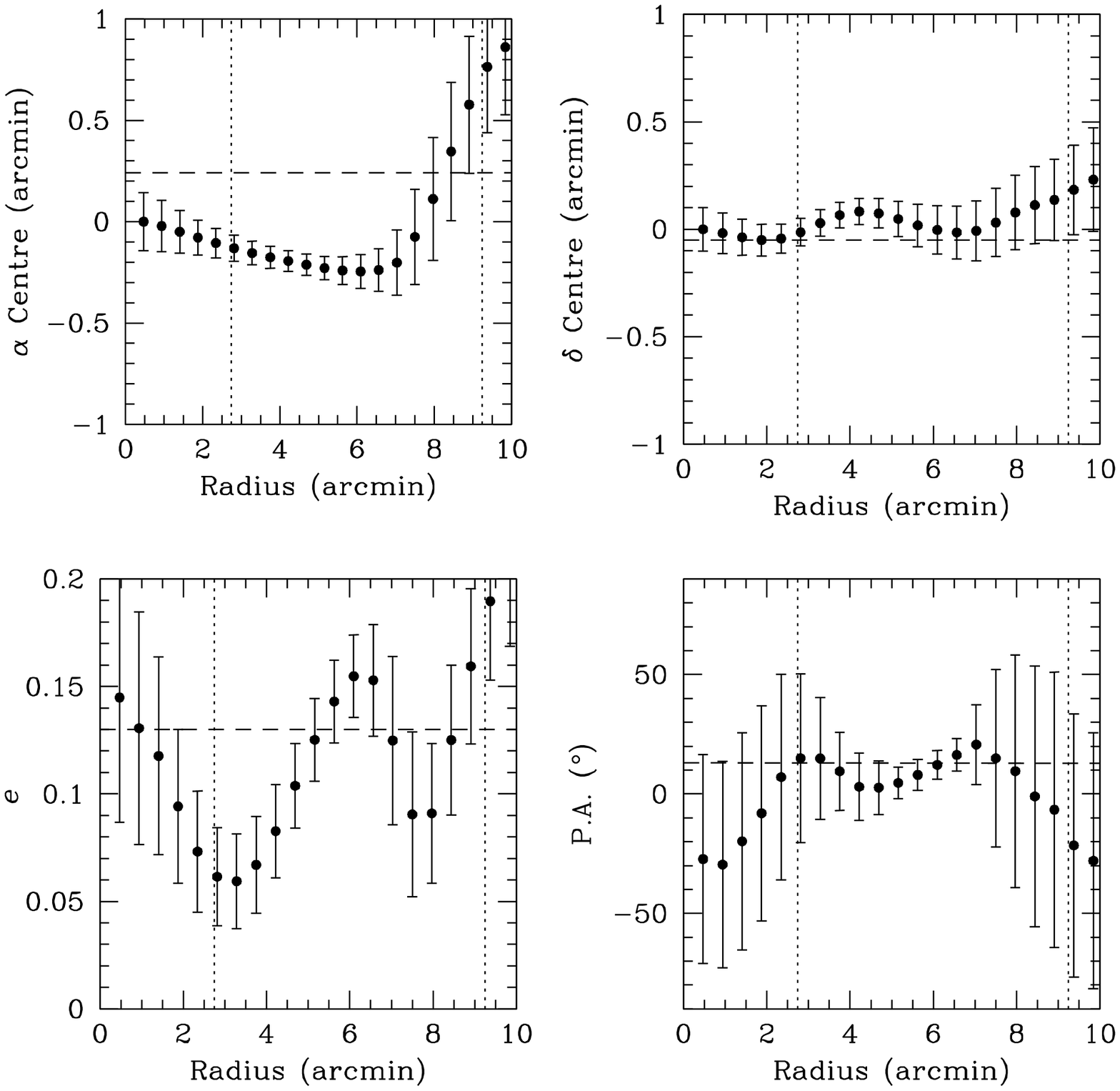}
\figcaption[Leo II internal structure]{The structure of Leo II as a function of major axis radius.  These datapoints were derived from the contour diagram in Fig.\ \ref{contour}, where we have calculated the best-fitting centre, ellipticity and position angle at succeeding radii.  The dashed lines represent the values listed by \citet{m98} and the dotted lines represent the core and tidal radii derived in this paper.  We have shifted the $y$-axis in the RA and Dec panels to place the first datapoint at zero.  \label{lscontour}}

\plotone{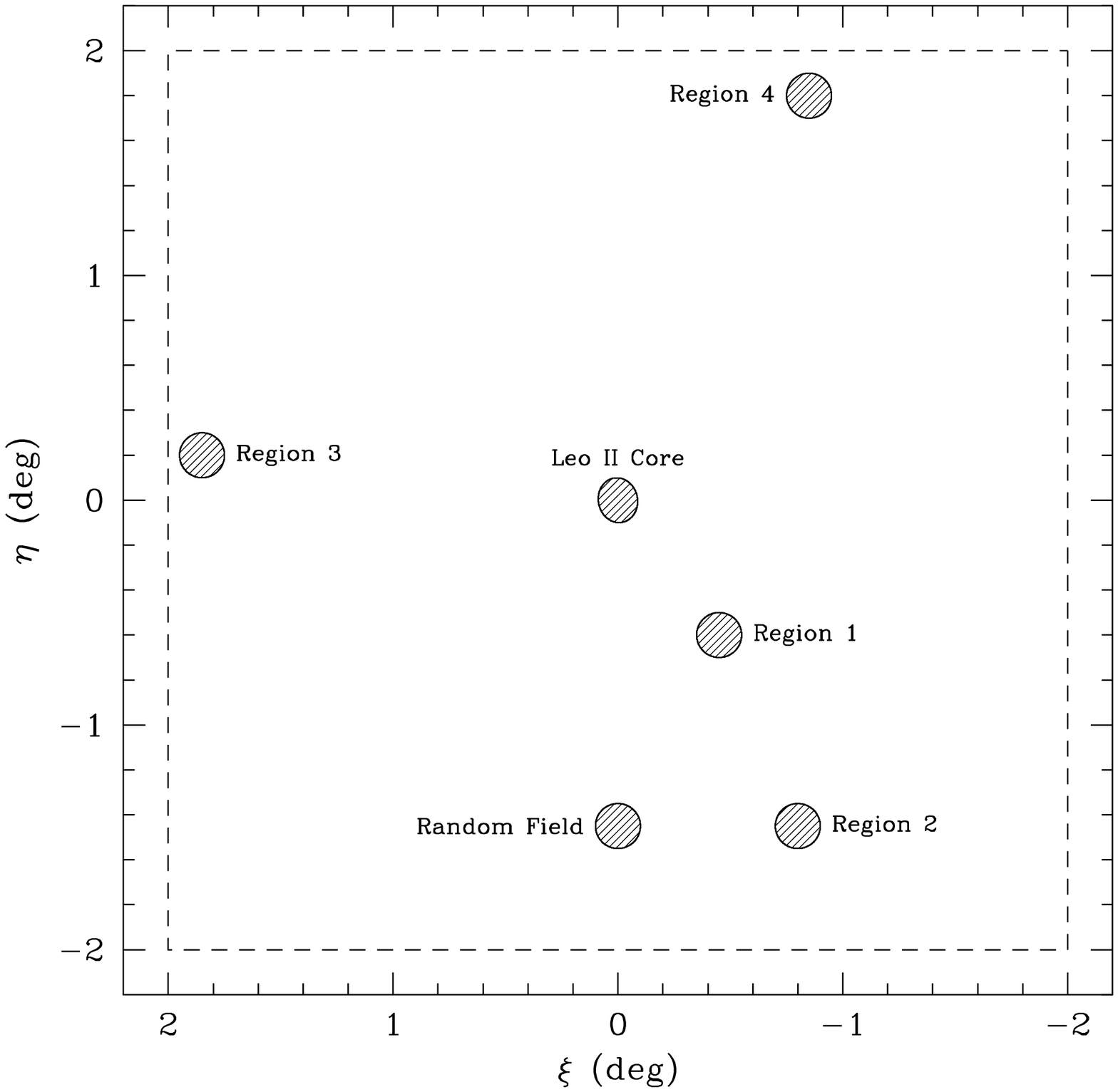}
\figcaption{Schematic diagram with the positions of the four overdensities and a random field region marked.  The dashed line represents the spatial limit of our survey.  \label{xyschem}}

\plotone{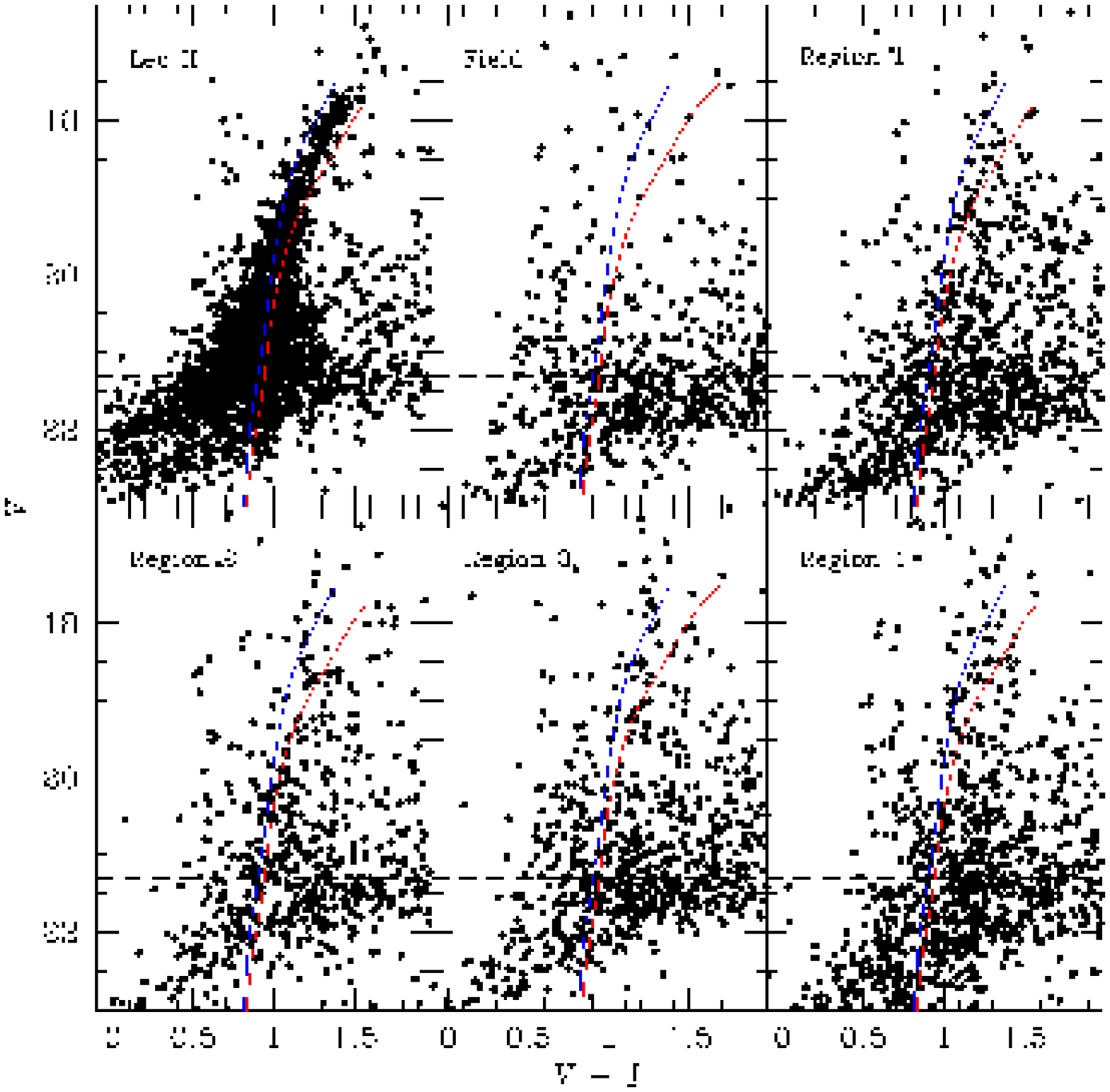}
\figcaption{CMDs for the central region of Leo II, a randomly selected field region, and the four overdense regions marked in Fig.\ \ref{xyschem}.  The isochrones represent a 12 Gyr population with [Fe/H] abundances of $-2.0$ (blue) and  $-1.5$ (red).  \label{clumpcmds}}

\clearpage

\plotone{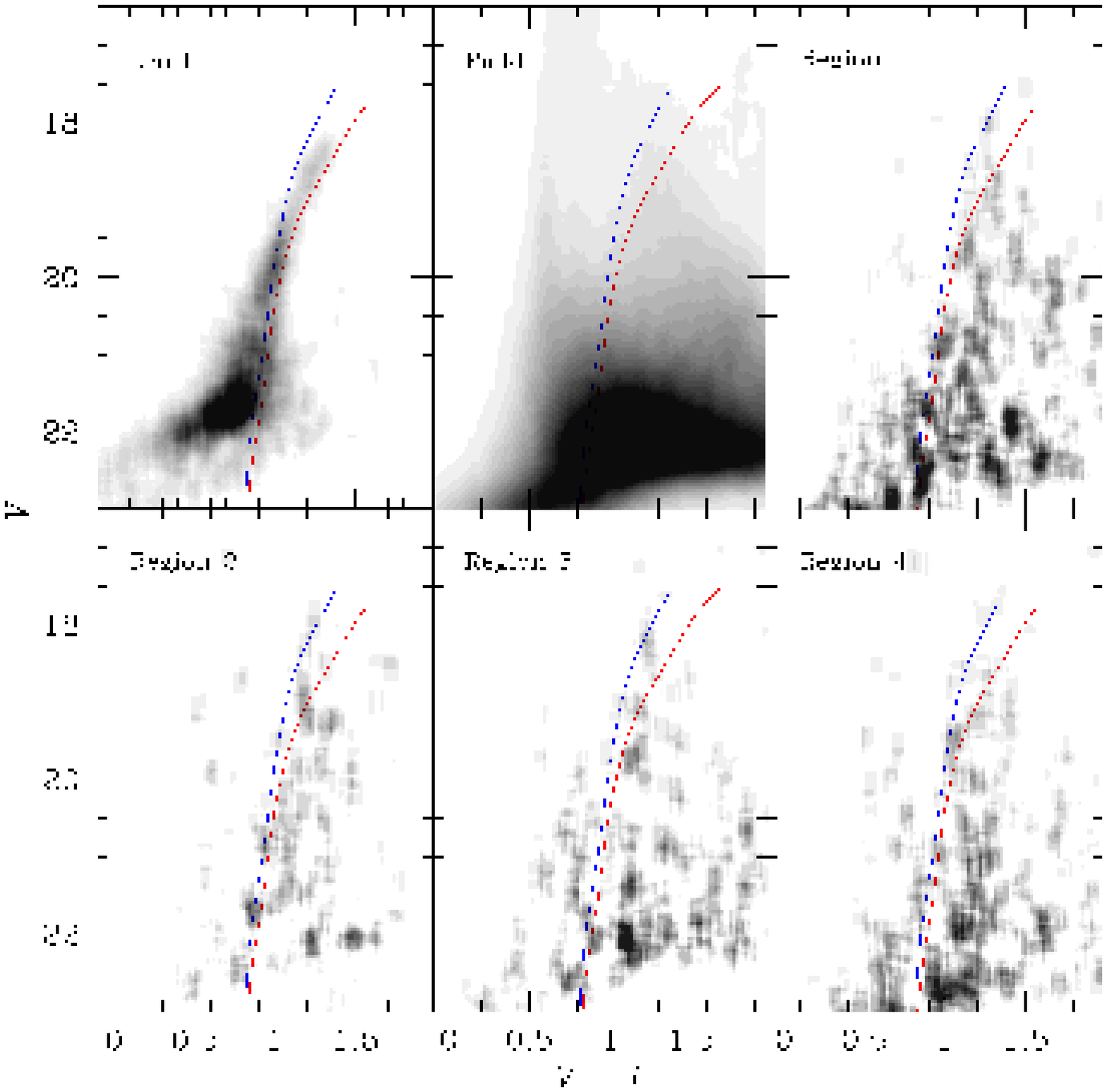}
\figcaption[Overdensity statistical subtraction]{Statistical subtraction of the field population from the extra-tidal stellar overdensities.  The Hess diagrams of the field region and the field-subtracted Leo II core are shown.  For comparison, the data for Region 1 represents approximately 300 stars.  Note the dissimilarity between the stellar populations of the overdensities compared to that of Leo II.  The two isochrones were derived for a 12 Gyr population with [Fe/H] abundances of $-2.0$ (blue) and  $-1.5$ (red).  \label{clumpcmdsub}}

\clearpage

\plotone{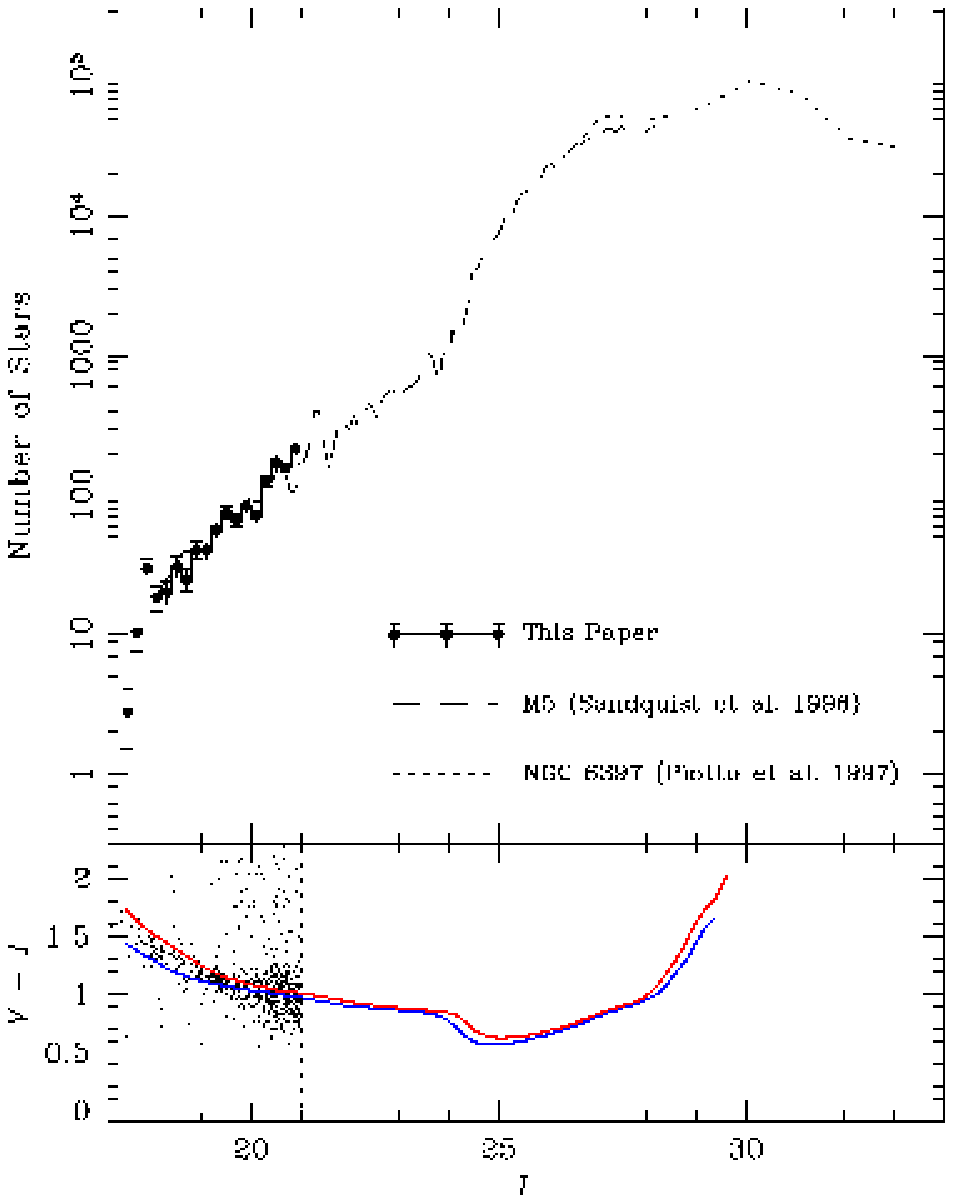}
\figcaption[Leo II luminosity function]{Luminosity function for Leo II.  The data (and Poisson uncertainties) from this paper are represented by the solid line.  We have extrapolated to fainter magnitudes using the luminosity functions for the globular clusters M5 (\citealt{sandquist96}; dashed line) and NGC 6397 (\citealt{piotto97}; dotted line) shifted by the appropriate distance moduli and stellar number.  The lower panel contains our Leo II data in Johnson-Cousins filters $V$ and $I$ (converted using the relations derived by \citealt{jordi06}) down to the completeness limit ($I=21.0$ is equivalent to $i=21.3$ at the centre of the Leo II RGB).  We have also placed 12 Gyr isochrones with [Fe/H] abundances of $-1.5$ (blue) and $-2.0$ (red) for reference.  \label{lf}}

\clearpage


\begin{table}
\centering
\caption[SDSS fields]{SDSS fields analysed with DAOPHOT}
\label{sdssfields}
\vspace{0.2cm}
\begin{tabular}{c|cccc}
\tableline
\tableline
Run & Rerun & Camcol & Fields & Section \\
\tableline
5183 & 4 & 40 & 245--247 & Leo II dSph \\
5183 & 5 & 40 & 245--247 & Leo II dSph \\
5194 & 5 & 40 & 360--362 & Leo II dSph \\
5140 & 1 & 40 & 112--114 & FN Leonis \\
5224 & 1 & 40 & 107--109 & FN Leonis \\
\tableline
\end{tabular}
\end{table}

\begin{table}
\begin{center}
\caption[Leo II Parameters]{Leo II Parameters}
\label{leoIIpars}
\vspace{0.2cm}
\begin{tabular}{l|cc}
\tableline
\tableline
Parameter & Value & Reference \\
\tableline
RA (J2000) & 11:13:28.8 & This paper \\
Dec (J2000) & 22:09:06.0 & This paper \\
Elliptcity, $e$ & 0.11 & This paper \\
Position angle & $6.7 \pm 0.9$ & This paper \\
$E(B-V)$ & $0.019 \pm 0.004$\tablenotemark{a} & \citet{schlegel98} \\
Distance & $233\pm 15$ kpc & \citet{bell05} \\
$(m - M)_0$ & $21.84 \pm 0.13$ & \citet{bell05} \\
 & & \\

$r_c$ & $2.64' \pm 0.19'$ & This paper \\
$r_t$ & $9.33' \pm 0.47'$ & This paper \\
$c=\log{(r_t/r_c)}$ & $0.55 \pm 0.05$ & This paper \\
$r_0$ & $2.2' \pm 0.1'$ & This paper \\
 & & \\
$\Sigma_{0,V}$ & $24.2 \pm 0.3$ mag/arcsec$^2$ & This paper \\
$\Sigma_{0,I}$ & $23.1 \pm 0.2$ mag/arcsec$^2$ & This paper \\
$L_V$ & $(7.4 \pm 2.0) \times 10^5 L_{\odot}$ & This paper \\
$L_I$ & $(11.4 \pm 3.0) \times 10^5 L_{\odot}$ & This paper \\
$M_V$ & $-9.9 \pm 0.3$ & This paper \\
$M_I$ & $-11.0 \pm 0.3$ & This paper \\
$(B-V)$ & $0.65 \pm 0.15$ & \citet{hodge82} \\
$(V-I)$ & $1.1 \pm 0.1$ & This paper \\
 & & \\
$v_r$ & $79.1 \pm 0.6$ km s$^{-1}$ & \citet{koch07c} \\
$\sigma_{v_r}$ & $6.6 \pm 0.7$ km s$^{-1}$ & \citet{koch07c} \\
$M_{\mbox{\scriptsize tot}}$ & $5.2^{+1.1}_{-1.0} \times 10^6 M_{\odot}$ / $6.6^{+1.5}_{-1.1} \times 10^7 M_{\odot}$ & This paper\tablenotemark{b} \\
$(M/L)_V$ & $7.0^{+1.5}_{-1.4}$ / $90^{+31}_{-30}$ & This paper\tablenotemark{b} \\
$(M/L)_I$ & $4.6^{+1.0}_{-0.9}$ / $58^{+20}_{-19}$ & This paper\tablenotemark{b} \\
\tableline
\end{tabular}
\end{center}
\tablenotetext{a}{The uncertainty quoted here represents the reddening variation over the $4 \times 4$ square degree field.}
\tablenotetext{b}{The first value was calculated under the `mass-follows-light' assumption, while the second adopted a dark matter halo with constant density throughout the stellar system and refers to the total mass within $r_t$}
\end{table}


\end{document}